\documentclass[aps,floats,superscriptaddress,pre,twocolumn]{revtex4}
\usepackage{amssymb,amsmath}
\usepackage{amsmath,amssymb}
\usepackage{graphicx}
\usepackage{psfrag,textcomp}

\def\beq{\begin{equation}}
\def\eeq{\end{equation}}
\def\bea{\begin{eqnarray}}
\def\eea{\end{eqnarray}}
\begin{document}
\title{Phase transitions and order in two-dimensional generalized nonlinear $\sigma$-models}
\author{Tirthankar Banerjee}\email{tirthankar.banerjee@saha.ac.in}
\affiliation{Condensed Matter Physics Division, Saha Institute of
Nuclear Physics, Calcutta 700064, India}
\author{Niladri Sarkar}\email{niladri.sarkar@saha.ac.in}
\affiliation{Condensed Matter Physics Division, Saha Institute of
Nuclear Physics, Calcutta 700064, India}
\author{Abhik Basu}\email{abhik.basu@saha.ac.in}
\affiliation{Condensed Matter Physics Division, Saha Institute of
Nuclear Physics, Calcutta 700064, India}
\date{\today}

\begin{abstract}
We study phase transitions and the nature of order in a class of 
classical generalized
$O(N)$ nonlinear  $\sigma$-models (NLS) constructed  by
minimally
  coupling pure NLS with additional degrees of freedom in the form of
  (i) Ising ferromagnetic spins, (ii) an advective Stokesian velocity and (iii)
  multiplicative noises.
  In  examples (i) and (ii), and also (iii) with the associated multiplicative 
noise
  being not sufficiently long-ranged, we show 
that the models may display a class of unusual phase transitions
between
  {\em stiff} and {\em soft phases}, where the effective spin stiffness,
respectively, diverges and vanishes in the long wavelength limit at two dimensions ($2d$),
unlike in pure NLS. In the stiff phase, in the thermodynamic limit
the variance of the transverse spin (or, the Goldstone mode)
  fluctuations are found to scale with the system size $L$ in $2d$ as $\ln\ln
L$ with a model-dependent amplitude, that is markedly
  weaker than the well-known $\ln L$-dependence of the variance of the broken
symmetry modes in
  models that display quasi-long range order in $2d$. Equivalently, for
$N=2$ at $2d$ the equal-time spin-spin correlations decay in powers of
inverse logarithm
of the
spatial separation with model-dependent exponents.
These
transitions are
controlled by the model
parameters those couple the $O(N)$ spins with the additional variables.
In the presence of long-range noises in example (iii), true long-range order 
may set in $2d$,
depending upon the specific details of the underlying dynamics.
  Our results should be useful  in understanding phase transitions in equilibrium and nonequilibrium
low-dimensional systems with continuous symmetries in general.

\end{abstract}

\maketitle

\section{Introduction}

Understanding phase transitions and order in low dimensional systems has 
remained a
topic of considerable theoretical interests. For two-dimensional ($2d$) systems
in
equilibrium with continuous symmetries and short-range interactions,
there is no phase transition to a low
temperature ($T$) broken symmetry phase with long range order (LRO) at
any $T\neq 0$. This is a consequence of the well-known Mermin-Wagner
theorem (MWT)~\cite{mermin}. As a result in the thermodynamic limit (TL),
there is no statistically flat phase of a fluid membrane or a $2d$
crystal  in thermal equilibrium.  In contrast, long-range
order in $2d$ equilibrium systems can exist at a finite $T$, if
there are effective long-range interactions present in the system
which can take the system out of the validity of the MWT. Notable
examples include tethered membranes~\cite{tethered}, where one finds a
finite-$T$ crumpling transition and a low-$T$
statistically flat phase with long-range orientational order~\cite{tethered} at
$2d$. Symmetric inhomogeneous fluid membranes form a closely related example, see Ref.~\cite{tirtha-mem1}. Here a phase 
transition is obtained
between a phase with a finite persistence length and a phase with a
diverging persistence length (or equivalently, a {\em stiffening
transition}), tuned by the coupling constant that couples the local
curvature with the local inhomogeneity order parameter. However, true LRO does 
not exist even in the phase with diverging persistence length in 
Ref.~\cite{tirtha-mem1}.
The latter phase is found to be characterized by a diverging bending
modulus in the long wavelength limit at $2d$.

The classical $O(N)$ nonlinear $\sigma$-model~\cite{stanley} (hereafter NLS)
has
found wide usage in a
variety of
topics, including condensed matter systems~\cite{condmat}, mathematical
physics~\cite{mathphys}, particle physics~\cite{partphys} and
cosmology~\cite{cosmo}. This is a paradigmatic model with continuous
symmetries~\cite{stanley,chaikin}, in which
the consequence of the MWT is succinctly visible in a low-$T$
expansion. At $d=2$ in equilibrium, one finds~\cite{chaikin} that
the model has no finite-$T$ ordered state with LRO. In this
article, we ask: what can introduce phase transitions and order in NLS at
$2d$? Taking cue from the examples of tethered membranes, we expect that the
predictions of MWT may be bypassed in the presence of long-range
interactions between the $O(N)$ spins. Since  pure NLS has only short-range
interactions between the
spins, additional degrees of freedom which may
create {\em effective} long-range interactions, are required to invalidate
the prediction of the MWT here. These additional fields can be interpreted as
representing spatio-temporally nontrivial environments coupled to NLS. The
model parameters defining pure NLS  now get modified or {\em renormalized}
by the coupled additional degrees of freedom.
Clearly, any putative LRO in a $2d$ generalized NLS should depend
upon the specific nature of the additional degrees of freedom and their
couplings to the $O(N)$ spins, i.e., the results should be model-dependent. Since we are interested
in a question of general principle, it is useful to study simple reduced models,
where explicit calculations may be done in straightforward ways.
To this end, we generalize NLS and
construct three different simple variants of NLS by
coupling it to additional fields through generic local interactions, one in
equilibrium and the
others out of equilibrium: We consider a  system of NLS, that is (i)
thermodynamically coupled with Ising spins - hereafter Model I, (ii)
dynamically coupled with a Stokesian velocity field that is affected by
the feedback from the $O(N)$ spins - hereafter Model II, and (iii) dynamically
coupled with multiplicative noises through symmetry-allowed minimal coupling -
hereafter Model III. For Model I and Model II, we consider only
 nonconserved dynamics while for Model III we consider both nonconserved and
conserved dynamics of NLS. In a {\em low noise variance} expansion that 
generalizes the 
low-$T$ expansion
for the
equilibrium NLS, we illustrate the
possibility of order in the NLS with additional fields at $2d$. Our principal
results are (i) both Model I and Model II and nonconserved Model III without
any long range noise have the lower critical dimension $d_L=2$; for $d>d_L$ all
these have a critical point separating a ``high noise'' paramagmetic phase
with short-range order (SRO) and a ``low noise'' ferromagnetic phase with LRO, 
(ii)
in $2d$, both
Model I and Model II, and the nonconserved version of Model III (without any
long-range noise) with
appropriate choices of the model parameters,
show phase transitions, tuned by the coupling constants, from a {\em disordered
soft}
phase with SRO, where the effective spin stiffness of the $O(N)$ spins vanishes 
over a
finite system size to a {\em stiff phase}, where it
 stiffens significantly, diverging as $\ln L$  for a system of linear
size $L$. As a result, in TL
the variance of the fluctuations of the transverse components of the spins
depend  on $L$
as $\ln\ln L$ at $2d$, a dependence
weaker than
the standard $\ln L$ dependence of
the variance of the elastic degree of freedom in an elastic
Hamiltonian that displays quasi long-range order (QLRO) in $2d$. For $N=2$ at
$2d$ this implies a spatial decay of the equal-time spin
correlations in powers of the logarithm of the spatial
separation with model-dependent exponents. The model-dependence of these
exponents are reminiscent of  the model-dependent exponents that characterizes 
the
well-known algebraic decay in QLRO; nevertheless, the spatial decay here is 
markedly slower
than the  algebraic decay in QLRO.  In contrast, the conserved version of
Model III without any long-range multiplicative noise admits no stiff phase at $2d$.
(ii) Nonconserved Model III with long-range multiplicative noises can display
true long-range order at $2d$, controlled by the noise amplitude and the
relevant
coupling constant. However, the conserved version of Model III with long-range
multiplicative noises does not display any LRO in $2d$. In addition, we
also calculate the dynamic exponent $z$ that characterizes the correlators of
the transverse spin fluctuations in the stiff phases at $2d$ and at the
unstable fixed point (FP) separating the  paramagnetic  and
 ferromagnetic phases for $d>2$.
The remainder of this article is organized as follows. In
Sec.~\ref{rev}, we provide a short review of NLS. In Sec.~\ref{one},
\ref{two} and \ref{model3}, we construct and
analyze our Model I, II, and III, respectively. In Sec.~\ref{conclu} we
summarize our
results. Some of the technical details are available in Appendices at the end.

\section{Short review on NLS}\label{rev}

The statistical mechanics of NLS with an
$N$-component spin $\Phi=(\phi_1,...,\phi_N)$ in the presence of an external 
magnetic field $h_i,\,i=1,..,N$ in $d$-dimensions is described by the free 
energy
functional ${\mathcal F}_{\sigma}$
~\cite{chaikin}
\begin{equation}
{\mathcal F}_\sigma = \frac{1}{2}\int d^dx [\kappa (\nabla_\alpha \Phi)^2
-h_i\phi_i],\label{sigmafree}
\end{equation}
where $\kappa$ is the spin stiffness; we
impose $\Phi^2=1$. By
means of a perturbative low-$T$ expansion,
it has been shown~\cite{chaikin} that the transition temperature $T_0$
separating the
low-$T$ ferromagnetic phase and the high-$T$ paramagnetic phase
is given by
\begin{equation}
 \frac{T_0}{\kappa}=\frac{2\pi\epsilon}{N-2}\label{nlstemp}
\end{equation}
at $d=2+\epsilon,\,\epsilon \geq 0$.
Thus, $T_0$ vanishes at $2d$ (i.e., $d_L=2$), precluding any finite $T$ ordered 
phase with
LRO. For $N=2$ at $2d$, $T_0$ is
indeterminate, suggesting
that degrees of freedom not included in NLS, e.g.,
topological
defects and amplitude fluctuations, should destroy the
order. Similar analyses have been done on the quantum version of the  
$O(N)$ nonlinear $\sigma$-model; see Ref.~\cite{uwe}.
We now ask whether or how these results may be significantly modified in the presence of
 additional degrees of freedom. We construct Models I, II and III as above to 
address
this issue. To set
 up the notations clearly, we denote the spin components of an $O(N)$ spin by 
Roman indices and the
components
of a vector or an operator in the real or Fourier space by Greek indices in the
remaining part of this article.

\section{Model I: NLS coupled with Ising spins}\label{one}

We study the equilibrium nonconserved relaxational dynamics of NLS coupled
with Ising spins. 
In the spirit of the Landau-Ginzburg coarse-grained approach, we
represent the Ising spins by a scalar field $\psi ({\bf x})$, where
$\bf x$ is the coordinate in a $d$-dimensional space. We start with
the combined free
 energy functional ${\mathcal F}_I$ for NLS coupled with the Ising spins:
Assuming a minimal symmetry-permitted
coupling between the Ising spins and the $O(N)$ spins, we write
\begin{eqnarray}\label{scalar_free}
 {\mathcal F}_I&=&\int d^{d}x[\frac{\kappa}{2}(\nabla_\alpha \Phi)^2-
{\bf h}\cdot \Phi+\frac{r}{2}\psi^2
 +\lambda(\nabla_\alpha
\Phi)^2\psi^2\nonumber
\\ &+&\frac{1}{2}(\nabla_\alpha\psi)^2+\frac{u}{4!}\psi^4],
\end{eqnarray}
where, as before, $ \Phi=(\phi_1,...,\phi_N)$, is an $N$-component
vector of unit modulus,i.e., $\Phi^2=1$.
Here, $r=T-T_c$, $T$ being the temperature, $T_c$ is the mean  field critical
temperature for the second order Ising magnetic transition, coupling constant
$u>0$. Setting $\psi=0$ yields Eq.~(\ref{sigmafree}) above; $\kappa$
is the spin stiffness. For $N=2$, (\ref{scalar_free}) is related to a
Ginzbug-Landau model recently proposed in studies on high temperature
superconductors~\cite{tvr}.  In 
(\ref{scalar_free}) the coupling between
$\phi_i$ and
$\psi$ are controlled by the coupling constant $\lambda$; With $T>T_c$, for all
$\lambda>0$, ${\mathcal F}_I$ is minimized by  uniform states of the fields,
i.e.,
$\phi_i=const.$ and $\psi=0$, where as, with large enough $\lambda<0$, such a
uniform state gets thermodynamically unstable. We choose $\lambda>0$ here. It
is constructed such that Eq.
(\ref{scalar_free}) is invariant under rotation of the $O(N)$ spin $\phi_i$ in
the spin space and inversion of the Ising spins $\psi\rightarrow -\psi$.
Usual order-disorder transitions and  relaxational dynamical critical behavior 
near critical points in a related model in higher dimensions (e.g., three dimensions)
have been studied in Ref.~\cite{folk}, where a variety of temperature driven 
phase transitions separating the disordered and the ordered phases are discussed. Unlike
Ref.~\cite{folk}, we here assume that the $O(N)$ spins are in the ordered state (and
hence put a fixed length constraint on the $O(N)$ spins) and
study the effects of the Ising spins coupled with the $O(N)$ spins via generic symmetry
allowed couplings on the putative ordered state of the $O(N)$ spins at $2d$. Since the Ising spins
undergo a usual paramagnetic to ferromagnetic transition at $T_c$, we focus on the state
of the assumed order of the $O(N)$ spins at $T=T_C$.

We are
interested in studying the purely relaxational nonconserved
dynamics. The equations of motion for $\Phi$ and $\psi$ are respectively given
by:
\begin{equation}\label{eom_scalar_phi}
 \frac{\partial\Phi}{\partial t}=-\Gamma\frac{\delta {\mathcal F}_I}{\delta
\Phi}+{\boldsymbol \theta}
 =\Gamma\left[\kappa\nabla^2\Phi+2\lambda\nabla_\beta
 (\psi^2\nabla_\beta\Phi)+{\bf h}\right] +{\boldsymbol\theta},
\end{equation}
and
\begin{eqnarray}\label{eom_scalar_psi}
 \frac{\partial\psi}{\partial t}&=&-\Gamma_2\frac{\delta {\mathcal F}_I}{\delta
\psi}+\eta\nonumber \\
 &=&\Gamma_2\left[-r\psi-2\lambda\psi(\nabla_\beta \Phi)^2+\nabla^2\psi +
\frac{u}{3!}\psi^3 \right]+\eta.
\end{eqnarray}
Here, $\Gamma$ and $\Gamma_2$ are the kinetic coefficients of $\Phi$ and $\psi$, 
respectively.
A quick inspection of Eq (\ref{eom_scalar_phi}) reveals that 
$\psi$-fluctuations contribute positively to $\kappa$ through the term with 
coefficient $\lambda$. Whether or not this can compete with the well-known {\em 
thermal softening} of $\kappa$ that is present even in pure NLS~\cite{chaikin} 
can only be determined after a thorough calculation. This remains a major goal 
of this work.
Stochastic functions
${\boldsymbol\theta}$ and $\eta$ are zero mean Gaussian white noises with
variances given by [we set
Boltzmann constant $k_B=1$]
\begin{eqnarray}
 \langle\theta_i(x,t)\theta_j(0,0)\rangle&=&2D\Gamma\delta_{ij}\delta({\bf
x})\delta(t),\label{thetavari}\\
 \langle\eta(x,t)\eta(0,0)\rangle&=&2\Gamma_2 T\delta({\bf x})\delta(t).
\end{eqnarray}
We are considering equilibrium dynamics of the coupled system. Hence,
the system obeys the Fluctuation-Dissipation Theorem (FDT)~\cite{chaikin,uwe}.
This enforces $D= T$.
Noting that $T/\kappa$ is a dimensionless number at $2d$, we intend to 
calculate the effective or {\em
renormalized} $\kappa$ in a low temperature expansion ($D/\kappa \ll 1$) up to
one-loop order. For simplicity we treat the fluctuations of $\psi$ up
to the harmonic order, i.e., we set $u=0$~\cite{comment2}.  For
simplicity, we now set $\Gamma_2=1$.

 It is clear that a uniform state of the $O(N)$ and Ising spins, where the
$O(N)$ and Ising spins are perfectly aligned among themselves, minimizes 
${\mathcal
F}_I$. Such a state may be parameterized as $\Phi=(1,0,...,0)$.  Thermal 
fluctuations should reduce the component of $\Phi$ along the
direction of order. Now assume an {\em ordered state} at low-$T$ where the
$O(N)$ spin
system is assumed to be ordered in a given direction. A
convenient parameterization of $\Phi$ for such a state is
$\Phi=(\sigma,{\boldsymbol\pi})$; thus, with
$\sigma\approx 1$ and  $\pi_i,i=1,..,N-1$ are the small transverse
fluctuations;  hence, $\boldsymbol \pi$ is an $N-1$ component vector. We write
$\sigma=\sqrt{1-\pi^2}
\approx1-\frac{\pi^2}{2}$, where $\pi^2 =\sum_i \pi_i^2$. We impose the
condition $\Phi^2=1$ on the dynamics via imposing
  a $\delta$-function, $\delta_1=\delta\left[\sigma^2+\pi^2-1\right]$ on the
  generating functional to be constructed (see below) from
(\ref{eom_scalar_phi}) and (\ref{eom_scalar_psi}). Notice that the fixed length
constraint $\Phi^2=1$ essentially implies $\Phi \cdot\partial_t\Phi=0$, i.e.,
the
variation in $\Phi$ takes place normal to $\Phi$, or, in other words, there are
$(N-1)$ independent transverse fluctuating degrees of freedom. The latter are
to be
associated with $\pi_i$ defined above with $\sigma\approx 1$. Thus, the
stochastic force ${\boldsymbol\theta}=\partial_t\Phi +\Gamma\delta {\mathcal
F}_I/\delta\Phi$ in (\ref{eom_scalar_phi}) must also be purely transverse.
This is ensured by imposing a second $\delta$-function
$\delta_2=\delta[\hat\Phi\cdot\Phi]$ on the associated generating
functional~\cite{uwe,bausch}, where 
$\hat\Phi=(\hat\sigma,\hat{\boldsymbol\pi})$ is
the dynamic conjugate of
$\Phi$~\cite{janssen}.
  Clearly, $\delta_1$ contributes only when $\sigma=\sqrt{1-\pi^2}$.
Further, with $\hat\Phi\cdot\Phi=\hat\sigma\hat\sigma+\hat{\boldsymbol
\pi}\cdot{\boldsymbol\pi}$,  $\delta_2$ contributes for
$\hat\sigma=-\frac{\hat\pi_i\pi_i}{\sigma}=-\frac{\hat\pi_i\pi_i}{\sqrt{1-\pi^2}
}$; here  $\hat \pi_i$ is the dynamic conjugate of $\pi_i$.

After inclusion of the constraints in the form of the $\delta$-functions
$\delta_1$ and $\delta_2$, the generating functional for the system is given by
(assume $\bf h$ is aligned along $\sigma$ with a magnitude $h_1$)
\begin{eqnarray}\label{gf1}
 {\mathcal Z}_I &=& \int {\mathcal D}\sigma {\mathcal D}\hat\sigma {\mathcal
D}\pi {\mathcal D}\hat\pi {\mathcal D}\psi{\mathcal
D}\hat\psi~{\exp[S_I]}\nonumber \\&=&\int
{\mathcal D}\sigma {\mathcal D}\hat\sigma {\mathcal D}\pi {\mathcal D}\hat\pi
{\mathcal D}\psi {\mathcal D}\hat\psi ~{\exp} [ \int d^dxdt \hat\sigma\hat\sigma
 \frac{D}{\Gamma}\nonumber \\&+& \int d^dxdt
\frac{D}{\Gamma}\hat\pi_j\hat\pi_j\nonumber \\&-&
 \int d^dxdt\hat\sigma \left\{\frac{1}{\Gamma}\frac{\partial\sigma}{\partial
t}-
\kappa\nabla^2\sigma-2\lambda\nabla_\beta(\psi^2\nabla_\beta\sigma)-h_1
\right\}\nonumber \\&-&
 \int d^dxdt \hat\pi_j \left\{\frac{1}{\Gamma}\frac{\partial\pi_j}{\partial t}-
\kappa\nabla^2\pi_j-2\lambda\nabla_\beta(\psi^2\nabla_\beta\pi_j)\right\}
\nonumber \\&+&
 \int d^dxdt T \hat\psi\hat\psi- \int d^dxdt \hat\psi \{\partial_t \psi +
 r\psi -\nabla^2\psi\nonumber \\
&+&2\lambda\psi(\nabla_{\beta}\sigma)^2+2\lambda\psi(\nabla_\beta\pi)^2\}]
\nonumber \\
&\times& \delta[\sigma^2+\pi^2-1]\delta[\hat\sigma\sigma+\hat\pi_i\pi_i]
\end{eqnarray}
The functional integrals over $\sigma$ and $\hat\sigma$ in ${\mathcal Z}_I$
above may be evaluated by using the two $\delta$-functions, which yield
 two Jacobians.
Overall the total Jacobian $J$ evaluates to ${(1-\pi^2)}$, appearing at every
$\bf x$ and $t$. Thus, the contribution of the total Jacobian
to $\mathcal Z$ is a product of all these terms,{\it viz.}~\cite{chaikin,janssen},
\begin{eqnarray}
 \Pi_{x,t}J&=&\Pi_{x,t} {(1-\pi^2)}=  {\exp}[-\sum_x \int {dt}~
{\ln}(1-\pi^2)]\nonumber \\
 &=&  {\exp}[-\rho \int {d^dxdt}~ {\ln}(1-\pi^2)],
\end{eqnarray}
where, $\rho$ is a number equal to the number of degrees of freedom per unit
volume that introduced
while going from summation to integration. We now expand the nonlinear terms up to
$O(T/\kappa)$, since we are interested in a low-$T$ one-loop expansion. We
obtain the action functional after truncating to the linear order
in $(T/\kappa)$
\begin{eqnarray}\label{action1}
 S_I &=& \int d^dxdt [\frac{D}{\Gamma}\hat\pi_j \hat\pi_j-\hat\pi_j
\{\frac{1}{\Gamma}\frac{\partial \pi_j}{\partial t}
-\kappa\nabla^2\pi_j\nonumber
\\&-&2\lambda\nabla_\beta(\psi^2\nabla_\beta\pi_j)\} +\frac { D } {
\Gamma}(\hat\pi_j\pi_j)^2\nonumber \\
 &+&\frac{1}{2}\hat\pi_j\pi_j \{ -\frac{1}{\Gamma}\frac{\partial \pi^2}{\partial
t}  +\kappa\nabla^2\pi^2 -2h_1(1+\pi^2/2) \}+
 T\hat\psi\hat\psi\nonumber \\&-& \hat\psi \{ \partial_t \psi+r\psi
-\nabla^2\psi+2\lambda\psi(\nabla_\beta \pi)^2 \} +\rho \pi^2]; \label{action-model1}
\end{eqnarray}
see Appendix~\ref{remodel1} for details.
Notice that while the contribution from $J$ nominally yields a quadratic term
of the form $\rho\pi^2$,
it is $O(T/\kappa)$~\cite{bausch1}; see also Appendix~\ref{remodel1}.
It is convenient to work in the Fourier space; we define $\bf q$ and
$\omega$ as the Fourier wavevector and frequency,
respectively. 
At $T>T_c$, the fluctuations in $\psi$ are short lived and small; as a result,
the contributions to the measurable physical quantities from the
$\psi$-fluctuations are small for $T>T_c$~\cite{lesstc}. In contrast, the
$\psi$-fluctuations diverge in the long wavelength limit as $T\rightarrow T_c$.
Similarly, $\pi_i$ being a broken symmetry mode (i.e., a Goldstone mode),
correlation $\langle |\pi_i ({\bf q},t)|^2\rangle$ diverges in the long
wavelength limit at all $T$. Thus, contributions to the measurable quantities
from $\psi$-fluctuations can compete with those originating from the broken
symmetry mode fluctuations only near the critical point, i.e., as $T\rightarrow
T_c$. Hence, in our subsequent analysis below in this Section, we set
$D=T=T_c$.
In order to deal with these long wavelength divergences in a systematic manner,
we employ Wilson momentum shell dynamic renormalization group
(DRG)~\cite{chaikin,halpin}; see also Ref.~\cite{uwe} for detailed 
discussions on DRG applications to dynamic critical phenomena. To this end, we 
first integrate out fields
$\pi_i({\bf q},\omega),\psi({\bf q},\omega),\hat\pi_i({\bf q},\omega),\hat\psi({\bf q},\omega)$ 
with wavevector
$\Lambda/b<q<\Lambda,\,b>1$, perturbatively up to the one-loop order in
(\ref{action1}). Here, $\Lambda$ is an upper cut off for wavevector. This allows
us to obtain the "new" model parameters corresponding to a modified action
$S_I^<$
with an upper cutoff $\Lambda/b<\Lambda$. We obtain
\begin{eqnarray}
\kappa^<&=&\kappa[1+\Delta+2\frac{\lambda}{\kappa}\tilde\Delta],\\
\left(\frac{D}{\Gamma}\right)^<&=&\frac{D}{\Gamma}[1+\Delta],\\
\frac{1}{\Gamma^<}&=&\frac{1}{\Gamma}[1+\Delta],\\
h_1^<&=&h_1[1+\frac{N-1}{2}\Delta].
\end{eqnarray}
 Here, superscript $^<$ refers to the parameters in the action with a  reduced
upper cut off $\Lambda/b$. Notice that $(D/\Gamma)^<$ and $1/\Gamma^<$ have the
same form, which is a consequence of the FDT. Furthermore,
  \begin{equation}
  \Delta=\int_{\Lambda/b}^\Lambda \frac{d^dq}{(2\pi)^d} \frac{D}{(\kappa
q^2 +
h_1)},\;\tilde\Delta=\int_{\Lambda/b}^\Lambda\frac{d^dq}{(2\pi)^d}\frac{T}{r+q^2
}
  \end{equation}
  where we have used the forms of the correlators given in (\ref{remodel2}).
  Clearly, at 2d, $\Delta$ is log divergent for small $h_1$ while $\tilde\Delta$
  is log divergent at $r(=T-T_c)=0$.

  In order to extract the renormalized parameters, we then rescale wavevectors
and frequencies according to ${\bf q}'=b{\bf q}$ and $\omega'=b^z{\bf \omega}$,
where $z$ is the dynamic exponent.
 Under these rescalings, fields $\phi_i,\hat\phi_i$ also scale. We write
$\hat\phi_i \rightarrow \hat\xi\hat\phi_i^{\prime}$ , $\phi_i \rightarrow
\xi\phi_i^{\prime}$
under the above rescalings. FDT in the present model
implies that $Im \langle \hat\pi_i\pi_i\rangle$ and
$\omega\langle\pi_i\pi_i\rangle$ must scale in the same way under the above
rescalings ($Im$ refers to the imaginary part). This consideration yields
$\hat\xi=\xi b^{-z}$. With this, rescaling factors for the various model
parameters may be obtained. These lead to the renormalized parameters
 \begin{eqnarray}\label{sc1 h}
        h_1^{\prime}&=&b^{-d-z}\hat\xi\xi h_1 \left[1+\frac{(N-1)}{2}\Delta \right],\\
     \kappa^{\prime}&=&\kappa \hat\xi\xi b^{-(d+z+2)} \left[ 1+ \Delta+
\frac{2\lambda}{\kappa}  \tilde\Delta \right],\label{kapp1ren}\\
     \left(\frac{D}{\Gamma}\right)^{\prime}&=&\hat\xi^2 b^{-(d+z)}\frac{D}{\Gamma} \left[ 1+\Delta \right],\label{fdt1}\\
    \left(\frac{1}{\Gamma}\right)^{\prime}&=&b^{-(d+2z)}\hat\xi\xi\frac{1}{\Gamma} \left[ 1+\Delta \right].\label{fdt2}
 \end{eqnarray}
 Notice that with $\hat\xi=\xi b^{-z}$, the left hand sides of Eqs.~(\ref{fdt1})
and (\ref{fdt2})  scale in the same way. This is a consequence of the FDT as
mentioned above.

In general, the  external magnetic field $\bf h$  itself is an $O(N)$ vector that couples
with the $O(N)$ spin $\Phi$ through the coupling with the external
magnetic field given by $-\int h_j\phi_j d^dx$ in the free energy.
Now consider the coupling
$-\int h_j\pi_j d^dx$ in the free energy (\ref{scalar_free})
which would produce a term $\int h_j\hat\pi_j d^dx dt$ in the corresponding action functional (\ref{action1}).
Requiring this "external part" of the action to be invariant under rescaling
implies $\int h_j\hat\pi_j \, d^dx dt
=\int h^{\prime}_j\hat\pi^{\prime}_j\, d^dx^{\prime} dt^{\prime}$. Since the
$O(N)$
symmetry of the problem with $h_i=0$ ensures that $\pi_i$ scales the same way as $\sigma$,
we have $h^{\prime}_i=\hat\xi h_i$; see Ref.~\cite{chaikin} for similar
arguments for pure NLS..
Due to the $O(N)$ symmetry of the model for $h_i=0$, all components of
$h_i$ must scale in the same manner. Hence,
\begin{equation}\label{sc2 h}
 h_1^{\prime}=\hat\xi h_1.
\end{equation}
Using Eq.~(\ref{sc1 h}) and Eq.~(\ref{sc2 h}), we now evaluate $\xi$ and $\hat\xi$.
\begin{eqnarray}
 \xi=b^{d+z} \left[1-\frac{(N-1)}{2}\Delta \right],\\
 \hat\xi=b^{d} \left[1-\frac{(N-1)}{2}\Delta \right]
\end{eqnarray}
These yield
\begin{equation}\label{corr_k}
 \kappa^{\prime}=\kappa b^{d-2} \left[ 1-(N-2)\Delta+\frac{2\lambda}{\kappa}\tilde\Delta \right].
\end{equation}
We set $D=T=T_c$ and are interested in the fluctuation corrections at
$d=2+\epsilon,\epsilon >0$.
At the leading order in $\epsilon$, we evaluate $\Delta$ and $\tilde\Delta$
above at $d=2$.
We thus find $\Delta=\frac{ T_c}{2\pi\kappa} {\ln}b$ for
small $h_1$ and $\tilde\Delta=\frac{T_c}{2\pi}{\ln}b$.

We now obtain the continuum flow equations for $\kappa$ and ${\Gamma}$. Let
$b={\rm e}^{l},\,l\rightarrow 0$ and $d=2+\epsilon$.
Using these definitions and Eq.~(\ref{corr_k}), the differential flow equation
for $\kappa$ becomes
\begin{equation}
 \frac{d\kappa}{dl}=\kappa \left[ \epsilon -\frac{(N-2) T_c}{2\pi\kappa}
 +\frac{2\lambda T_c}{2\pi\kappa} \right].\label{kflow1}
\end{equation}
First consider $N>2$. For $N-2>2\lambda$, there is a
DRG fixed point (FP)
given by $d\kappa/dl=0$. This yields at the unstable FP
\begin{equation}
\frac{T_c}{\kappa}=\frac{2\pi\epsilon}{(N-2)-2\lambda},\label{tcsigma1}
\end{equation}
at $d=2+\epsilon$. Compare (\ref{tcsigma1}) with the corresponding  result for
pure NLS for the
(reduced) transition
temperature given by (\ref{nlstemp}). Thus,
(\ref{tcsigma1})  implies a (reduced) transition temperature 
$\tilde T=T_c/\kappa$ having a value $\tilde T^*\sim O(\epsilon)$ at the FP 
that separates a low
temperature ferromagnetic phase
(ordered
state), where the majority of spins are aligned  and a
high temperature paramagnetic
phase without any alignment of spins at $d=2+\epsilon$. Since $\tilde
T^*=0$ at $2d$ ($\epsilon=0)$, we have $d_L=2$. This physical
picture holds for
$N>2+2\lambda$ for which $T^*$ drops to zero at $2d$. At $N=2+2\lambda$, $T^*$
is indeterminate at $2d$. This is the analog of the behavior of the transition
temperature for pure NLS with $N=2$. It is
understood that topological defects, e.g., vortices and amplitude fluctuations
destroy the order for $N=2$, $d=2$ in pure NLS~\cite{chaikin}. Since $N$ must 
be an
integer, $N$ may be increased in steps of unity, where as $\lambda$ is any real
number. Thus, $\lambda=1/2$ yields $N=3$ such that at $d=2$, $T^*$ is
indeterminate for Model I with $N=3$. It will be interesting to
study what role, if any, topological defects play to destroy the order there.

 In the above analysis, we have assumed the existence of an unstable FP,
 which may be ensured by choosing $\lambda$, a free parameter in the model,
appropriately. For $N < 2+2\lambda$, $d\kappa/dl >0$ thus precluding any FP. To
proceed further, we define
 a critical $\lambda_c(\epsilon)$, at which $\frac{d\kappa}{d l}$ vanishes, as a function of the $\epsilon$ parameter given by
\begin{equation}
 \lambda_c(\epsilon)=  \frac{(N-2)}{2} - \frac{\pi\kappa\epsilon}{T_c}.\label{critlam}
\end{equation}
Defining $\Delta\lambda(\epsilon)=\lambda-\lambda_c(\epsilon)$, (\ref{kflow1}) at $d=2$ may be written as
\begin{equation}\label{kappa_rl}
\frac{d\kappa}{dl}=\frac{T_c\Delta\lambda(\epsilon=0)}{\pi},
\end{equation}
giving an unstable FP $\lambda=\lambda_c(\epsilon=0)=(N-2)/2$ at $2d$. This is
reminiscent of a usual equilibrium second order phase transition; nonetheless, 
there are significant differences that we elaborate below.

 Consider now the scale-dependent or renormalized $\kappa(l)$ at an
off-critical point ($\lambda\neq \lambda_c$ or $\Delta\lambda \neq 0$).
  Clearly from Eq.~(\ref{kappa_rl}), for $\Delta\lambda(\epsilon=0) <0$,
  $\kappa(l)$ reduces linearly with $l$ at $2d$ and ultimately
 becomes zero at a particular length scale depending on $\kappa_0$ and
$\lambda$.  Thus even for NLS
 coupled to  Ising spins at $T_c$, with $\lambda < \lambda_c(\epsilon=0)$,
 $\kappa$ vanishes for a sufficiently large system at $2d$.
The discrete
 recursion relation for $\kappa$ at 2$d$ is given by
 \begin{equation}
  \kappa=\kappa_0 -\frac{(N-2) T_c}{2\pi} {\ln}b +\frac{\lambda T_c}{\pi}
{\ln}b,
 \end{equation}
 where $\kappa_0$ is a microscopic stiffness.
 Further writing ${\ln}b={\rm ln}(\Lambda/q)=-{\ln}(q a_0)$, where $a_0 \sim$
a microscopic
 length, we obtain the $q$-dependence of $\kappa$ :
 \begin{equation}
  \kappa(q)=\kappa_0 +\frac{(N-2) T_c}{2\pi}{\ln}(q a_0)-\frac{\lambda T_c}{\pi}{\ln}(q a_0).\label{kappq}
 \end{equation}
 Evidently, the second and third terms of (\ref{kappq}) reveal the 
competition between thermal softening of $\kappa$ and its stiffening due to the
$\psi$-fluctuations.
 Not surprisingly, the wavevector-dependent effective or renormalized $\kappa$
does not depend upon $\Gamma$,
a kinetic coefficient. This is again a consequence of the FDT-obeying dynamics.
 Equation~(\ref{kappq}) allows us to define a persistence length $\zeta \sim 1/q$, such that
 $\kappa(\zeta)=0$ (see, e.g., \cite{tirtha-mem1,peliti} for similar
 persistence lengths in different problems). This yields
 \begin{equation}
  \zeta=a_0 {\exp} \left[ \frac{2\pi\kappa_0}{(N-2)T_c -2\lambda T_c} \right].\label{persis}
 \end{equation}
 Hence, as $\lambda \rightarrow \lambda_{c_-}(\epsilon=0)=\frac{(N-2)}{2}$ in $2d$, the persistence
 length, $\zeta \rightarrow \infty$. Given the physical interpretation that
$\zeta$ is a typical length scale
at which $\kappa(\zeta)=0$, the system remains ordered over a length scale
smaller than $\zeta$.  Thus for system sizes larger than $\zeta$, there will 
only be SRO. We also define persistence length $\zeta(NLS)$ for pure NLS
($\lambda=0$)
\begin{equation}
 \zeta(NLS)=a_0 {\exp} \left[ \frac{2\pi\kappa_0}{(N-2)T_c}
\right].\label{zetanls}
\end{equation}
Clearly, even for $\lambda<\lambda_c$, $\zeta(NLS)<\zeta$. Thus, even in the
soft phase, the persistence length is larger than that for pure NLS, implying
that $\lambda$ in general favors order.

\begin{figure}[htb]
\includegraphics[height=6cm]{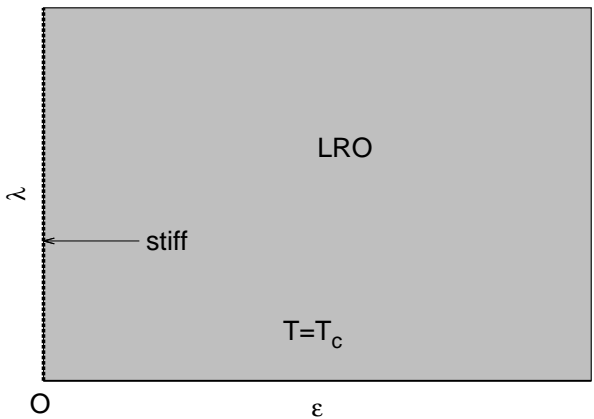}
 \caption{ Phase diagram in the $\epsilon-\lambda$ plane for $N=2$ 
($\lambda_c=0$). The stiff phase exists along the $\lambda$-axis (broken line, 
$\epsilon=0,\lambda>0$). Symbol $O$ which marks the origin $(0,0)$, corresponds 
to pure NLS with SRO at $2d$.}\label{phase1}
\end{figure}

 In contrast for $\lambda > \lambda_c(\epsilon=0)$, $\kappa$ grows on successive
applications of momentum shell DRG.
 Thus, at larger scales NLS is supposed to appear {\em more ordered}
than what it is at
 smaller scales. Clearly at higher dimensions, as $\epsilon$ increases,
 $\lambda_c(\epsilon)$ decreases. Hence at higher dimensions, weaker $O(N)$-Ising couplings are
 enough for $\kappa(l)$ to grow. Consider now the explicit form for  $\kappa(q)$
in the limit $q\rightarrow 0$ for $\lambda>\lambda_c(\epsilon=0)$ at $2d$.
For $\lambda
-\lambda_c(\epsilon=0)=\Delta \lambda (\epsilon=0)>0$ at $2d$,
 we have,
 \begin{equation}
  \kappa(q)=-\frac{T_c\Delta \lambda(\epsilon=0) }{\pi} {\ln}(q a_0) + \kappa_0,
 \end{equation}
 which diverges logarithmically in TL ($q\rightarrow 0$).
 In order to comment on the nature of order, if any, displayed by Model I, we 
look at the variance of the transverse spin fluctuations $\pi$,
given by $\Delta_0=\langle |\pi_i({\bf q},t)|^2\rangle$, now defined in terms of
the renormalized bending modulus $\kappa(q)$. We find
\begin{eqnarray}
 \Delta_0 &=& \int \frac{d^2 q}{(2\pi)^2} \frac{ T_c}{\kappa(q) q^2 + h_1}
 = -\frac{1}{2\Delta \lambda(\epsilon=0)} \int \frac{d q}{q {\ln}(q
a_0)}\nonumber \\
 &=& \frac{1}{2\Delta \lambda(\epsilon=0)}{\ln} |{\ln}
(a_0/L)|=\frac{1}{2\Delta\lambda(\epsilon=0)}\ln\ln
L,\nonumber \\\label{fluctuations_pi}
\end{eqnarray}
 in the limit
of large systems, i.e., $L\gg a_0$, where $L$ is a lower momentum cut-off 
$\sim$
linear system size.
Regardless of the value of $T_c$, (\ref{fluctuations_pi}) holds as long as
$\lambda > \lambda_c(\epsilon=0)$. Thus similar to Ref.~\cite{tirtha-mem1},
$\Delta_0$ diverges as $L \rightarrow \infty$, but very slowly
and remains finite for any finite value of $L$.  In contrast
for  pure NLS, $\kappa(q)$ vanishes for a low
enough $q$. Hence for the pure system, $\Delta_0$ will diverge even for
a finite system size. For $d>2$, it may be shown straightforwardly that
$\Delta_0$ is finite in TL. Thus, true LRO exists for $d>2$. A phase diagram
of the system in the $\lambda-\epsilon$ plane is shown in Fig.~\ref{phase1},
with the line of FPs given by $\lambda=\lambda_c(\epsilon)$, SRO 
for
$\lambda<\lambda_c(\epsilon)$, LRO for
$\lambda>\lambda_c(\epsilon),\,\epsilon>0$ and stiff phase for
$\lambda>\lambda_c(\epsilon=0)$ .

It is insightful to consider the flow of the {\em reduced temperature}
$\tilde T=T_c/\kappa$ ar $2d$. We find
\begin{equation}
\frac{d\tilde T}{dl}=\tilde T^2[\frac{N-2}{2\pi}-\frac{2\lambda}{2\pi}]. \label{tflow}
\end{equation}
For $\lambda=0$ and $N>2$, $\tilde T=0$ is the only DRG FP. This is, however,
an unstable FP (expected), such that even for an arbitrarily small
(microscopic) $\tilde T= \tilde T_0$,  scale-dependent $\tilde T(l)$ grows.
Integrating (\ref{tflow})
\begin{equation}
 -\frac{1}{\tilde T}=A_0 l + C,
\end{equation}
where $A_0=\frac{N-2}{2\pi}-\frac{2\lambda}{2\pi}$ and $C$ is a constant of integration. Assume $A_0>0$.
Set at $l=0$, $1/\tilde T=1/\tilde T_0 >0$, we
have
\begin{equation}
 -\frac{1}{\tilde T}=A_0 l-\frac{1}{\tilde T_0}.
\end{equation}
Thus, as $l \rightarrow 1/(A_0\tilde T_0)\sim \ln\zeta$ from below, $\tilde
T\rightarrow \infty$.
This implies that for any microscopic $\tilde T_0>0$, the large
scale properties of the system is identical to that of a system at $\tilde
T\rightarrow\infty$, indicating disordered phase at all $\tilde T>0$. Since
$T_c$ for an Ising system at $2d$ is always larger than zero (there is a
finite temperature Ising ferromagnetic phase), our Model I at $T=T_c$ is always
in
its disordered phase for $\lambda=0$.  This scenario changes
for $A_0<0$, or, $\lambda >\lambda_c$ at $2d$. Flow equation (\ref{tflow}), upon integration,
now yields
\begin{equation}
 \frac{1}{\tilde T}=|A_0|l+\frac{1}{\tilde T_0}.
\end{equation}
Evidently, for $l\rightarrow\infty$, $\tilde T\rightarrow 0$.
Therefore, for $\lambda>\lambda_c(\epsilon=0)$,
scale-dependent $\tilde T(l)$ flows to zero as $l\rightarrow\infty$,
where as for $\lambda<\lambda_c(\epsilon=0)$,
$\tilde T(l)$ flows to infinity as $l\rightarrow \ln \zeta$. Thus for $\lambda>\lambda_c(\epsilon=0)$,  the
properties of the system at the largest (formally TL) scale is expected
to be same as that of the system at zero temperature. This suggests the 
existence of
order in the system. On the other hand, for $\lambda<\lambda_c(\epsilon=0)$, the behavior of
the system at large scale is same as that at infinite temperature, indicating a
paramagnetic phase with SRO only. Thus, as $\lambda$ rises from 0 
to a high value through
$\lambda_c(\epsilon=0)$, the system undergoes a phase transition from a disordered
(paramagnetic) phase to an ordered phase through a critical
point at $\lambda=\lambda_c(\epsilon=0)$. This is reminiscent of the usual temperature
driven order-disorder second order transition; however, in the presence case,
$T$ is kept fixed at $T_c$ and $\lambda$ is the tuning parameter.

\begin{figure}[htb]
 \includegraphics[height=7cm]{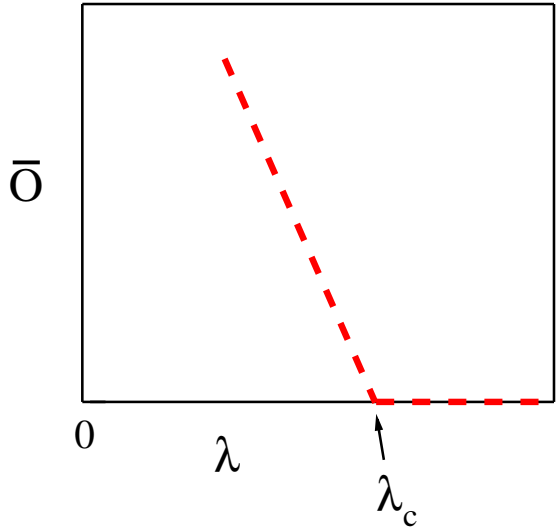}
 \caption{(color online)Schematic variation of the order parameter $\overline O$ versus
$\lambda$ in $2d$ for $N>2$.}
 \label{orderparam}
\end{figure}

\begin{figure}[htb]
 \includegraphics[height=6.5cm]{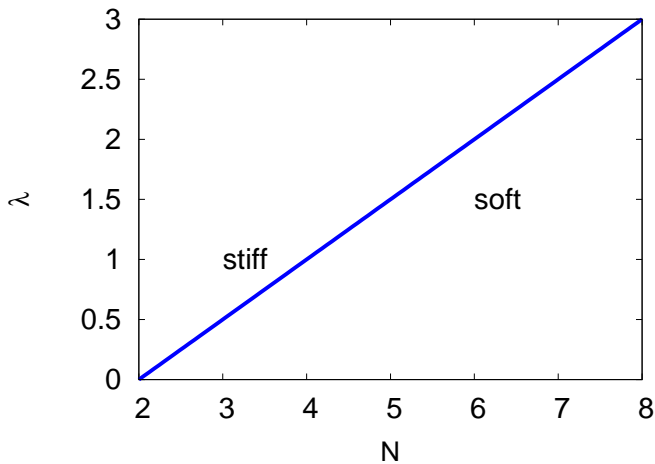}
 \caption{(color online) Schematic phase diagram in $N-\lambda$ plane at $2d$ at $T=T_c$. The
thick line (blue)
corresponds to the equation $\lambda_c(\epsilon=0)=(N-2)/2$. The stiff and soft phases are
marked (see text).}
 \label{lamN}
\end{figure}

\begin{figure}[htb]
 \includegraphics[height=6.5cm]{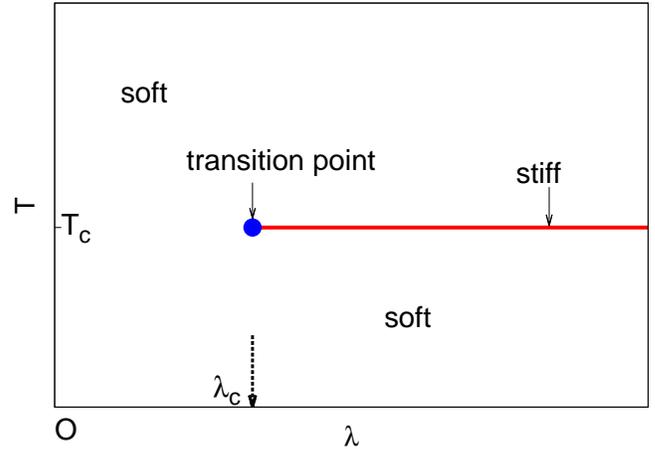}
 \caption{(color online) Schematic phase diagram is $\lambda-T$ plane for $N>2$ at $2d$.
 Symbol, $O=(0,0)$ is the origin. Stiff phases exist
along the horizontal thick line (red). Thick circle (blue) refers to the critical point
$(\lambda_c,T_c)$.}
 \label{Tlam}
\end{figure}

We now discuss the nature
of the order in the ordered phase, and whether it is identical to the
magnetic (i.e., ferromagnetic) order in usual magnetic systems. The usual magnetic transition
is conveniently described by the magnetic order parameter, which is non-zero in the ferromagnetic
phase but zero in the paramagnetic phase. Since we assume order in the ``1'' direction in the
spin space, we take $m=\langle\sigma\rangle=\langle\sqrt{1-\pi^2}\rangle\approx
1-\langle\pi^2\rangle/2$ as the order parameter. Clearly, in the disordered phase
$m\approx 0$ for a system with a linear size approaching $\zeta$. 
Interestingly,
even in the ordered phase, $m\approx 0$ in TL, due to the
divergences
found in $\Delta_0$ above; see Eq.~(\ref{fluctuations_pi}). Thus, $m$ remains zero on {\em both}
sides of the critical point given by $\lambda=\lambda_c$, and hence cannot work 
as 
an order parameter here~\cite{foot1}. Instead, we take
\begin{equation}
\overline O = [\ln
(\xi/a_0)]^{-1}=-T_c\Delta\lambda/(\pi\kappa_0),\label{op}
\end{equation}
with $\lambda<\lambda_c$ and
$\overline O=0$ with $\lambda\geq \lambda_c$ as the order
parameter at 2d. This is similar to the order parameter defined in the context of a
crumpled to stiff phase of a heterogeneous fluid membrane~\cite{tirtha-mem1}.
Evidently, $\overline O$ in the soft phase  rises smoothly from zero,
as $\lambda$ is reduced from $\lambda_c$. Thus, with
$\lambda$ as the control parameter, the {\em order parameter exponent}
is unity. In the stiff phase, $\overline O$ is naturally zero. A schematic plot
of $\overline O$ versus $\lambda$ is shown in Fig.~\ref{orderparam}.
Fig.~\ref{lamN} and Fig.~\ref{Tlam}  provide  schematic phase diagrams of
Model I in the $N-\lambda$ (with $T=T_c$) and $\lambda-T$ (for a fixed $N>2$)
planes at $2d$~\cite{foot2}.

What happens for $N=2$ at $2d$? In pure NLS, the
transition
temperature becomes indeterminate, suggesting the importance of topological
defects and amplitude fluctuations in destroying LRO in the
system. When the $O(2)$ spin is coupled to the Ising spin, the flow equation for
$\kappa(l)$ reduces to
\begin{equation}
\frac{d\kappa}{dl}=\frac{\lambda T_c}{\pi},
\end{equation}
giving $\kappa(q)=\kappa_0-\lambda T_c\ln(a_0q)/\pi$ for all $\lambda>0$. Thus
for {\em any positive} $\lambda$,
renormalized $\kappa(l)$ diverges in TL. As a result,
$\Delta_0\sim [\ln\,\ln L]/\lambda$ at $T=T_c$ and the $O(2)$ spins
should appear more ordered than the XY model with QLRO at low $T$.
In an equivalent picture, (\ref{tflow}) for $N=2$ reduces to
\beq
\frac{d\tilde T}{dl}=-\tilde T^2\frac{\lambda}{\pi}.
\label{tflow_2}
\end{equation}
Thus, $\tilde T(l)$ {\em always} flows to zero for all $\lambda>0$. Hence,
Model I with $N=2$ is in
its ordered stiff phase for
 {\em all}
$\lambda$. We now calculate the Debye-Waller factor and the equal-time
spin-spin correlation
 $C_s(r)=\langle \Phi({\bf x})\cdot\Phi({\bf x'})\rangle$,
for $N=2$ in the stiff phase  ($T=T_c, \lambda>\lambda_c$),
where $r=|{\bf x}-{\bf x'}|$ is assumed to be large. For $N=2$, the
transverse components $\boldsymbol\pi$ has only one component, say,
$\tilde\pi$;
$\Phi$ may be expressed as a complex number $\exp (i\Psi)$ where $\Psi({\bf
x},t)$ is a real scalar field; we write $\Phi_1=\cos\Psi,\,\Phi_2=\sin\Psi$
as the two components of $\Phi$. For small transverse fluctuations in a nearly
ordered state, associate $\Psi$ with $\tilde\pi$ (equivalently, parametrize
$\Phi=\exp (i\tilde\pi)$; see Appendix~\ref{O2} for formulation of the
dynamics in terms of $\Psi$). In the renormalized version of Model I, the
variance of $\tilde\pi ({\bf x},t)$ in the stiff phase (for all $\lambda>0$)
in $2d$ is given by (\ref{fluctuations_pi}) with $\epsilon=0$ and $N=2$. Now
assuming ``1'' to be the ordering direction,
 the average order parameter
\begin{eqnarray}
 \langle\cos\Psi\rangle &=&\exp[-\frac{1}{2}\langle \tilde\pi({\bf
x},t)^2\rangle]=\exp[-\frac{1}{2}\Delta_0]\nonumber \\&=&\exp[-\frac{1}{4
\lambda}\ln |\ln L|]=\exp
(-W),
\end{eqnarray}
in (renormalized) Model I for a system of (large) linear size $L$.
The Debye-Waller factor that determines the depreciation of the magnetization
due to the noise induced fluctuations (thermal fluctuations for equilibrium
systems) is given by $\exp (-2W)$, where  $2W=\Delta_0=(\ln \ln
L)/(2\lambda)$ in the limit of large $L$~\cite{chaikin}. Clearly,
$W\rightarrow\infty$ in TL here and hence
the order parameter vanishes in TL. Note, however, that as $\lambda$ rises, $W$
decreases for a fixed $L$, and thus, an increase in $\lambda$ reduces thermal
depreciation of the magnetisation.
Furthermore,
\begin{eqnarray}
 C_s(r)&=&\langle\cos [{\Psi({\bf x},t)-\Psi(0,t)}]\rangle \nonumber \\&=& Re
\langle\exp (i[\Psi({\bf x},t)-\Psi(0,t)])\rangle\nonumber =\exp[-g({\bf
x})],
\end{eqnarray}
where $g({\bf x})=\langle [\tilde {\pi}({\bf x},t)-\tilde\pi(
0,t)]^2\rangle/2 = \int \frac{d^2 q}{(2\pi)^2} [\langle |\tilde\pi_{\bf
q}(t)|^2\rangle [1-\exp (i{\bf q}\cdot{\bf x})]$, where we again have made the
association between the phase field $\Phi$ and $\tilde\pi$ for small transverse
fluctuations. Defining ${\bf q\cdot x}=qr\cos\varphi=u\cos\varphi$, we obtain
in renormalized Model I
\begin{eqnarray}
 g({\bf x})&=&-\frac{1}{2\lambda}\int_{rL^{-1}}^{r\Lambda}
du\frac{1-J_0 (u)}{u\ln (ua_0/r)}\nonumber \\ &=&
-\frac{1}{2\lambda}\int_{rL^{-1}}^1
du\frac{1-J_0(u)}{u\ln(ua_0/r)}+\frac{1}{2\lambda}\int_1^{
r\Lambda } \frac { du J_0(u)}{u\ln (ua_0/r)} \nonumber \\ &-&
\frac{1}{2\lambda}\int_1^{r\Lambda}\frac{du}{u\ln (ua_0/r)},\label{gx}
\end{eqnarray}
where $\Lambda\sim 1/a_0$ is an upper momentum cutoff; $J_0(u)$ is the zeroth
order Bessel function of first kind. Clearly, the first two terms in the last
line of (\ref{gx}) remain finite in TL, $L\rightarrow\infty$. Thus in TL for
$r\rightarrow\infty$ and neglecting the finite parts
\begin{equation}
 g({\bf x})=\frac{1}{2\lambda}\ln|\ln(a_0/r)|. \label{qlro}
\end{equation}
This gives
\begin{eqnarray}
 C_s(r)&=&\exp[-\frac{1}{2\lambda}\ln|\ln
(a_0/r)|]=\frac{1}{|\ln(a_0/r)|^{1/2\lambda}}\nonumber \\&=&\frac{1}{(\ln
r)^{1/2\lambda}},\label{qlro1}
\end{eqnarray}
in the limit $r\rightarrow\infty$.
Thus, the spatial dependence of $C_s(r)$ is characterized by a model-dependent
exponent $\lambda$, which is reminiscent of QLRO in the XY model below the
Kosterlitz-Thouless (KT) transition. Nonetheless, the spatial dependence in
(\ref{qlro1}) is markedly weaker than the algebraic decay of equal-time
correlators associated with QLRO~\cite{chaikin}. Again for a larger $\lambda$,
$g({\bf x})$ is smaller and the spatial decay of $C_s(r)$ is accordingly
weaker. This is consistent with the role of $\lambda$ in favoring order in the
system.

We now evaluate the dynamic critical exponent $z$ of the broken symmetry
modes $\pi_i$. The flow equation for $\kappa\Gamma$ is obtained as
\begin{equation}
 \frac{d}{dl}\kappa\Gamma = \kappa\frac{d\Gamma}{dl}+\Gamma\frac{d\kappa}{dl}.
\end{equation}
The last term vanishes at $\lambda\leq\lambda_c (\epsilon)$ yielding
\begin{equation}
 \frac{d}{dl}\kappa\Gamma = -\kappa\Gamma \left[ 2+\epsilon-z-
\frac{(N-2)T_c}{2\pi\kappa} \right].
\end{equation}
At the FP, this yields $z= 2+\epsilon-\frac{(N-2)T_c}{2\pi\kappa}$. When
$\lambda \leq \lambda_c(\epsilon)$, $T_c/\kappa=2\pi\epsilon/[N-2-2\lambda]$ at the DRG FP, which gives
$z=2-\frac{2\lambda\epsilon}{(N-2)-2\lambda}$ at the DRG FP. Thus $z$ clearly 
depends on $\lambda$ and decreases as $\lambda$ rises for a given $N$ and 
$\epsilon$. On the other hand, for $\lambda >\lambda_c (\epsilon)$ and using
(\ref{kflow1}), we find $z=2$, since $\frac{T_c}{\kappa(L)} \rightarrow 0$ for
$L\rightarrow\infty$ in the stiff phase. Notice that the dynamic exponent $z_\psi$ for $\psi$
with a nonconserved relaxational dynamics is $z_\psi=2$  to the linear order
in $u$. In addition, at $2d$ $z=2$ at the unstable FP as well. Thus, $z_\psi=z$ 
at $2d$ in general and strong dynamic scaling prevails.


\section{Model II: NLS coupled with a Stokesian velocity
field}\label{two}

Next,
we  construct a simple  nonequilibrium extension of NLS by coupling it to a
velocity field $\bf v$  dynamically
via advection. This may be relevant, e.g., in the dynamics of a collection 
of
$2d$ nematic liquid crystals in a $2d$ flow. These are
represented by a unit vector in the coarse-grained limit~\cite{degennes-book}, 
similar to the $O(N)$ spins (with $N=2$). Of course, the dynamics of 
equilibrium or {\em active}~\cite{sriram-RMP} nematics are far more complex 
than our simplistic model here; nonetheless, our results in this Section may 
serve as a prototype for the orientational dynamics in systems coupled with an 
advective velocity field.
The usual relaxational equation of motion for $\Phi$ is then
supplemented by an advective nonlinearity. We obtain
\begin{equation}\label{EOM for v}
 \partial_t \Phi + \tilde \lambda  {\bf v}. {\boldsymbol\nabla} \Phi= -\Gamma
\frac{\delta {\mathcal F}_\sigma}{\delta \Phi} + {\boldsymbol\theta},
\end{equation}
where $\tilde\lambda$ is a coupling constant.
For simplicity we assume $\bf v$ to obey a stochastically forced generalized
Stokes equation (neglecting inertia) of the form
\begin{equation}
 \eta \nabla^2 v_\alpha -\nabla_\alpha \Pi=\alpha_0{\boldsymbol
\nabla_\alpha}\Phi\cdot\nabla^2\Phi +f_\alpha.\label{veq}
\end{equation}
 The first term on the right hand side of (\ref{veq}) is a symmetry-allowed
 feedback of the $O(N)$ spin on $\bf v$. Such a feedback may originate
from a stress tensor due to the $O(N)$ spins
$\Sigma_{\alpha\beta}=-\alpha_0{\boldsymbol\nabla}_\alpha\Phi\cdot
{\boldsymbol\nabla}_\beta\Phi$, such that the stress vanishes for
$\Phi=const.$. This feedback term in (\ref{veq}) is reminiscent of an
$N$-component generalization of the feedback term in the generalized Navier
Stokes equation for a non-critical binary fluid mixture~\cite{binary}. 
We allow the
coupling constant $\alpha_0$ to be both positive or negative. Further $\Pi$ is
the generalized pressure. Notice that the coupling of $\Phi$ with $v_\alpha$
is of {\em dynamic origin}, unlike in Model I, where such couplings are of
static or thermodynamic origin (i.e., can be obtained from a free energy 
functional). Model II
reduces to pure NLS for $\bf v=0$.
Stochastic force
$\bf f$ is assumed to be zero-mean with a Gaussian distribution, having a
variance in the Fourier space that is given by
\begin{equation}
\langle f_\alpha({\bf q},t) f_{\beta}(-{\bf q},0) \rangle=D_0q^2 \delta (t)
\delta_{\alpha\beta}.
\end{equation}
Noise ${\boldsymbol\theta}$ is again chosen to be a Gaussian zero mean white noise with a
variance given by (\ref{thetavari}). Note however that being out of
equilibrium, $D$ in (\ref{thetavari}) no longer has the interpretation of the
temperature, although $D$ still has the same physical dimension as $T$. Thus,
$D/\kappa$ continues to be a dimensionless number. In what follows below, we
generalize the usual low $T$ expansion of equilibrium NLS and expand in small
$D/\kappa$. Apart from the formal similarity between $D$ here and $T$, it is
reasonable to expect that a low noise, marked by a low value of $D$, should
favor setting any  order; in contrast, a high value of $D$ (i.e., high noise)
should
distablize it.
Thus, an expansion in $D/\kappa$ is not only a formal generalization
of the standard low $T$ expansion, it is also physically meaningful.
We enforce incompressibility on $\bf v$. Thus, $\Pi$ may be
eliminated by imposing ${\boldsymbol\nabla}\cdot {\bf v}=0$ on (\ref{veq}).
We eliminate $\Pi$ to express $\bf v$  as
\begin{equation}
v_\alpha = \frac{\alpha_0 P_{\alpha\beta}}{\eta\nabla^2}(\nabla_\beta
\Phi\cdot\nabla^2\Phi)
+ \frac{P_{\alpha\beta}}{\eta\nabla^2}f_\beta.\label{finalv}
\end{equation}
Here,
$P_{\alpha\beta}=\delta_{\alpha\beta}-\partial_\alpha\partial_\beta/\nabla^2$ is
the transverse projection operator.
Equation (\ref{finalv}) may be used to eliminate $\bf v$ in (\ref{EOM for v}).
Evidently,
$v_\alpha$ enters into the dynamics of $\Phi$ in two distinctly different
ways: through (i) the multiplicative noise $f_\alpha$ and (ii) the
deterministic
term with the coupling constant $\alpha_0$.
Now proceeding as for Model I and replacing $\bf v$ by (\ref{finalv}), 
we obtain the action functional
\begin{eqnarray}\label{action_2}
 S_{II} &=& \int d^dxdt [\frac{D}{\Gamma}\hat\pi_i\hat\pi_i-\hat\pi_i \{\frac{1}{\Gamma}\frac{\partial \pi_i}{\partial t}
 -\kappa\nabla^2\pi_i\}+\frac{D}{\Gamma}(\hat\pi_i\pi_i)^2\nonumber \\
 &+&\frac{1}{2}\hat\pi_i\pi_i \{ -\frac{1}{\Gamma}\frac{\partial \pi_j^2}{\partial t}
 +\kappa\nabla^2\pi_j^2 -2h_1(1+\pi^2/2) \}]\nonumber \\
 &-& \frac{\tilde\lambda\alpha_0}{\Gamma}\int d^dx dt \hat \pi_i
[\frac{P_{\alpha\beta}}{\eta\nabla^2}(\nabla_\beta\pi_j)(\nabla^2 \pi_j)]
\nabla_\alpha \pi_i\nonumber \\&-&\int d^dx dt \lambda \hat\pi_i
\frac{P_{\alpha\beta}}{\eta\nabla^2}f_\beta\nabla_\alpha\pi_j,\label{action2}
\end{eqnarray}
by expanding  in (assumed
small)  $D/\kappa$, akin to  the Model I above in Sec.~\ref{one}; $\lambda=\tilde \lambda/\Gamma$;
$\pi_i,\,i=1,..,N-1$ is the component of the spin fluctuations transverse to
the
ordering direction and is a $(N-1)$-component vector.  Notice that there is no
analog of
$T_c$ of Model I in the present study; thus, unlike Model I, we do not restrict
ourselves to any particular value of $D$ (which is the nonequilibrium
analog of $T$ here). As
before, in order to proceed systematically, we use Wilson momentum
shell DRG to evaluate the loop integrals. Due to the couplings $\lambda$ and
$\alpha_0$, there are additional corrections to $\kappa$ over and above
the corrections in the pure NLS model. In
particular, $\kappa$ receives a
$O(\alpha_0 \lambda)$ correction at the lowest order in $D/\kappa$.
We find
 \begin{eqnarray}
 \left(\frac{D}{\Gamma} \right)^{<}&=&\frac{D}{\Gamma}+\frac{D}{\Gamma}\Delta,
 \\
  \left(\frac{1}{\Gamma} \right)^{<}&=&\frac{1}{\Gamma}(1+\Delta),\\
  \kappa^{<}&=&\kappa[1+\Delta (1-\mu)],\\
    h_1^{<}&=&h_1+\frac{h_1(N-1)}{2} \Delta,
 \end{eqnarray}
where, $\Delta=\int_{\Lambda/b}^\Lambda \frac{d^d q}{(2\pi)^d}\frac{D}{\kappa 
q^2+h_1}$, same as in
Model I and $\mu=\tilde\lambda \alpha_0/(\kappa\eta\Gamma)$; superscript $<$ 
has the same
implication as in our analysis of Model I above. Since
$\Delta\sim D/\kappa$, the above one-loop corrections are already $O(D/\kappa)$.
Thus, we need to find out only $O(D/\kappa)^0$ corrections, if any, to
$\lambda$
and $\alpha_0$. We show in Appendix~\ref{noren} that there are indeed no
corrections
to $\lambda$ and $\alpha_0$ at $O(D/\kappa)$.
 To evaluate the corrections we once again employ Wilson momentum DRG. We scale
the fields, $q$, $\Omega$, $\hat\phi_i$ and $\phi_i$
 in the same way as for Model I.
  The parameters are thus scaled in the following way:
 \begin{eqnarray}
 h_1^{\prime}&=& \hat\xi \xi b^{-d-z}h_1[1+\frac{N-1}{2} \Delta],\\
    \left(\frac{D}{\Gamma} \right)^{\prime}&=&\hat\xi^2 b^{-d-z}
\frac{D}{\Gamma}[1+\Delta],\label{2_DG_cor},\\
    \left(\frac{1}{\Gamma} \right)^{\prime}&=&\hat\xi\xi b^{-(d+2z)}
\frac{1}{\Gamma}[1+\Delta],\label{2_G_cor}\\
   \kappa^{\prime}&=&\hat\xi\xi b^{-(d+2+z)}\kappa[1+\Delta
(1-\mu)].\label{2_k_cor}
 \end{eqnarray}
Substituting Eq.~(\ref{2_G_cor}) in Eq.~(\ref{2_DG_cor}), we obtain,
\begin{equation}\label{D and Dprime}
 D^{\prime}=\hat\xi\xi^{-1}Db^z.
\end{equation}
 In Model I above, we have argued that
$Im\langle\hat\pi_m({\bf q},\omega) \pi_m({\bf -q},-\omega)\rangle$ and
$\omega\langle|\pi_m ({\bf q},\omega)|^2\rangle$
must scale in the same way due to FDT. We used this to set $D'=D$. Model II
does not satisfy FDT. Nonetheless,
using the freedom to choose one of the rescaling factors ($\hat\xi$ or $\xi$)
freely, we continue to impose $D^{\prime}=D$.
Thus, from Eq.~(\ref{D and Dprime}), $\hat\xi=\xi b^{-z}$. Once again
arguing that all components of $h_i$ should
scale in the same manner, $\xi$ evaluates to $b^{d+z}[1-\frac{N-1}{2}\Delta]$.
Hence, $\hat\xi=b^d[1-\frac{N-1}{2}\Delta]$.
\newline
Using these values of $\hat\xi$ and $\xi$ in Eq.~(\ref{2_k_cor}), we get
\begin{equation}
 \kappa^{\prime}=\kappa b^{d-2}[1-(N-2)\Delta -\mu\Delta].
\end{equation}
We substitute $b=e^{l},\,l\rightarrow 0$ and obtain
\begin{equation}
 \frac{d\kappa}{dl}=\kappa[\epsilon-\frac{(N-2)D}{2\pi\kappa}-\frac{\mu
D}{2\pi\kappa}].\label{kflow2}
\end{equation}
 The physical interpretation of (\ref{kflow2})
follows our analysis of (\ref{kflow1}) in Model I closely. As long as
$N>2-\mu$
there exists a DRG FP for the flow equation (\ref{kflow2}), such
that at the FP we have for the reduced noise strength $\tilde D=D/\kappa$
\begin{equation}
\tilde D^*=\frac{2\pi\epsilon}{N-2+\mu},\label{dkapp2}
\end{equation}
yielding a (reduced) critical noise strength  $\tilde D^*\sim O(\epsilon)$,
demarcating between a
ferromagnetic phase (ordered state) at  $\tilde D<\tilde
D^*$
and a
paramagnetic  phase (disordered state) at  $\tilde D>\tilde D^*$
at
$d=2+\epsilon$. Equation (\ref{dkapp2}) clearly shows $d_L=2$ for Model II.
Since the minimum physically realisable value of $N$ is 2,
$N>2-\mu$ for $\mu>0$ is always satisfied and hence a DRG FP always
exists.
Thus for $N>2-\mu$, $\tilde D^*=0$ at $2d$, lifting the indeterminacy of the
transition for $N=2$ NLS at $2d$. What this implies about the role of
topological defects at $2d$ for $N=2$
requires further investigations. Since in general, $\tilde D^*=0$ at $2d$ for
$\mu >0$ and $N\geq 2$, there are no ordered phase at $2d$. This is similar to
the results for pure NLS at $2d$.

\begin{figure}[htb]
 \includegraphics[width=9cm,height=6.5cm]{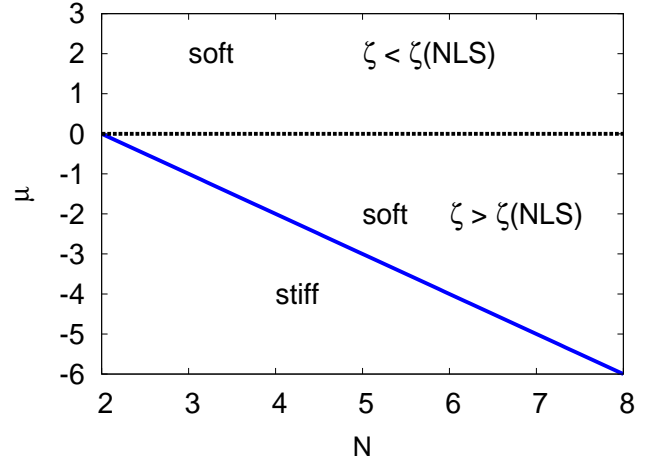}
 \caption{(color online)Schematic phase diagram of Model II in $N-\mu$ plane at $2d$. The
solid inclined (blue) line corresponds to the equation $N=2-\mu$, giving $\mu_c$. Soft
and stiff phases are marked (see text).}
 \label{muN}
\end{figure}

\begin{figure}[htb]
 \includegraphics[width=9cm]{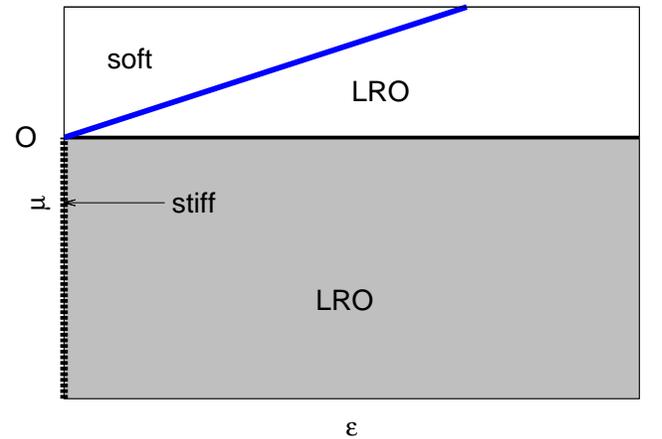}
 \caption{(color online)Schematic phase diagram of Model II in $\epsilon-\mu$ plane for $N=2$. Symbol, $O$
 marks the origin, $(0,0)$. $\mu_c=0$ for $\epsilon=0$. The negative $\mu$-axis (broken) gives the stiff phase
 for $\epsilon=0$. The inclined line (blue) corresponds to the equation $\mu=\frac{2\pi\kappa \epsilon}{D}$
 which separates phases with SRO (soft) and LRO for $\mu, \epsilon >0$.}
 \label{phase2}
\end{figure}

Consider now $\mu<0$. The DRG FP of (\ref{kflow2}) ceases to
exist for
$N<2+|\mu|$. Let us focus at $2d$. We define a critical $\mu_c$ by
$|\mu_c|=N-2,\,\mu_c <0$, such that at $2d$, $d\kappa/dl=0$ yields an unstable
FP given by $\mu=\mu_c$
identically [see (\ref{kflow2}) above]. This again is reminiscent of a second 
order
transition at $\mu=\mu_c$; see discussions for Model I above. Note that for
$N=2,\,\mu_c=0$. For
$\mu<\mu_c$, we again obtain a renormalized $q$-dependent
stiffness $\kappa(q) =-|\mu|D\ln (a_0 q)/(2\pi)$ at $2d$, as in Model I above.
Naturally, this implies for the variance $\Delta_0=( \ln\ln
L)/|\mu|$ in $2d$ for large $L$, again similar to Model I. Thus, as $\mu$ is
varied, Model II
undergoes a phase transition between a soft phase with a vanishing effective
stiffness $\kappa_e$ for a large enough system and a stiff phase with a
diverging $\kappa_e$ in TL. Nonetheless, this transition
for Model II is purely a nonequilibrium transition, due to the nonequilibrium
origin of the coupling $\mu$. In contrast, the corresponding transition in
Model I is an equilibrium phase transition. One can analogously define a
persistence length $\zeta$ such that $\kappa(\zeta)=0$, such that $\zeta$
remains finite in the disordered phase but diverges as $\mu\rightarrow\mu_c
<0$ at $2d$. As in Model I, the stiff phase for $\mu<\mu_c$ at $2d$ has no true
LRO; the magnetic order parameter $m=\langle\sigma\rangle$ vanishes
for systems in TL both in the ordered and the disordered phases at $2d$, and
hence, cannot be used to describe the transition at $2d$. Again like Model I,
order parameter defined by (\ref{op}) may be used to delineate the ordered phase
from the disordered one. The {\em order parameter exponent}, unsurprisingly, is
1, same as in Model I. Notice that for $\mu>0$, $\zeta <\zeta (NLS)$;
hence, an increasing $\mu$ makes the system more
disordered. The flow equation for the scale-dependent reduced noise
strength $\tilde D$ may be calculated in analogy with the flow equation for
$\tilde T$ in Model I; the associated physical interpretations for the
solutions of $\tilde D(l)$ are similar to those for $\tilde T(l)$ in Model I.
Specifically at $2d$ for $N=2$,
\begin{equation}
 \frac{d\tilde D}{dl}= \tilde D^2\frac{\mu}{2\pi}.
\end{equation}
Thus, for any $\mu <0$, $\tilde D (l)$ {\em always} flows to zero
corresponding to the stiff phase, where for any $\mu >0$, $\tilde D(l)$
diverges as $l\rightarrow \ln\zeta$, corresponding to the soft phase. This is
consistent with $\mu_c=0$ for $N=2$. The Debye-Waller factor $\exp(-2W)$ and
the
spin-spin correlation function $C_s(r)$ for $N=2$ at $2d$ may be calculated in
analogy with Model
I. Not surprisingly, $W\rightarrow \infty$ at $2d$ in the stiff phase
($\mu<0$), demonstrating the absence of LRO in the stiff phase. In addition,
$C_s(r)\sim [\ln r]^{-1/|\mu|}$ for $\mu<0$ and $r\rightarrow\infty$ at
$2d$. This is similar to the form of $C_s(r)$ in Model I stiff phase at $2d$,
with the model-dependent exponent being determined by $\mu<0$. See 
Fig.~\ref{muN} for a schematic
phase diagram of Model II in the $N-\mu$ plane at $2d$, highlighting the soft
and stiff phases. A corresponding phase diagram of
Model II in the $\epsilon-\mu$ plane is shown in Fig.~\ref{phase2}.


We now evaluate the dynamical critical exponent, $z$.
Using $b=e^{l}, d=2+\epsilon$ and Eq.~\ref{2_G_cor}, we obtain the
flow equation for $\kappa\Gamma$
\begin{eqnarray}
 \frac{d(\Gamma\kappa)}{dl}&=&\kappa\Gamma \left[z-2-\epsilon + \frac{(N-2) 
D}{2 \pi \kappa}
\right] \nonumber \\ 
&+&\kappa\Gamma \left[\epsilon-\frac{(N-2)D}{2\pi\kappa}-\frac{\mu
D}{2\pi\kappa}\right].
\end{eqnarray}
At the DRG FP, $d\kappa/dl=0$ and hence, therefore, $z=2+\epsilon-\frac{(N-2) 
D}{2 \pi \kappa}$, as long
as the FP exists. Since for $\mu>\mu_c$, $D/\kappa=2\pi\epsilon/(N-2+\mu)$, we
have $z=2+\frac{\mu\epsilon}{N-2+\mu}$ that depends upon $\mu$ at the DRG FP 
(similar to the $\lambda$-dependence of $z$ at the DRG FP in Model I). On
the other hand for $\mu < \mu_c$, we have 
 $z=2$ in the stiff phase.

In our above analysis, we have treated  $\mu$ as a {\em
constant} or {\em number}. This is justifiable provided it can be
shown that $\mu$ is indeed {\em marginal} under the spatial
and temporal rescaling mentioned above. We now show that below. Note
that there are no corrections to  $\mu$ at $O(D/\kappa)^0$;
see Appendix~\ref{noren}. Thus, $\mu$ is to be affected
only trivially under the rescaling of space and time. Consider the term
in Eq.~(\ref{action_2}), $\lambda \int d^dx dt \hat\pi_m {\bf
v}.{\boldsymbol\nabla}\pi_m$. In the Fourier space after rescaling
of momenta, frequencies and fields, this term scales as
\begin{eqnarray} \label{lam_sc}
&&\lambda b^{-2d-2z-1} \hat\xi\xi\xi_v\int d^dq_1 d^dq_2 d\Omega_1
d\Omega_2 \hat\pi_m({\bf -q}_1-{\bf q}_2,\nonumber \\ &-&\Omega_1-\Omega_2)
 \times v_\beta({\bf q}_2,\Omega_2) iq_{1\beta} \pi_m ({\bf q}_1,\Omega_1).
 \end{eqnarray}
This yields,
\begin{equation}\label{2_lam}
 \lambda^{\prime}=\lambda b^{-2d-2z-1} \hat\xi\xi\xi_v,
\end{equation}
where $\xi_v$ is the corresponding rescaling factor for
$v_\beta({\bf q},\Omega)$. Under rescaling $q^{\prime}=b q$ and
$\Omega^{\prime}=b^z \Omega$, we can write,
\begin{equation}
 \xi_f^2 \langle f_\alpha ({\bf q}_1',\Omega_1')f_\beta ({\bf
q}_2',\Omega_2')\rangle = 2D_0q_1^2 \delta ({\bf q}_1'+{\bf q}_2')\delta
(\Omega_1^{\prime} +\Omega_2^{\prime})b^{d+z-2},
\end{equation}
yielding $\xi_f=b^{(d+z-2)/2}$, where $\xi_f$ is the rescaling factor for the
noise $f_\alpha$. Now apply the same rescaling of wavevectors and frequencies
on Eq.~(\ref{finalv}), yielding in the Fourier space~\cite{convolution}
\begin{eqnarray}\label{conv_scal}
 \xi_v v_\alpha ({\bf q}',\Omega^{\prime})&=& b^{-(d+z+1)}\alpha_0\xi^2
\left[\frac{P_{\alpha\beta}}{\eta\nabla^2}(\nabla_\beta
\pi_j)\nabla^2\pi_j\right]_{{\bf
q}',\Omega'}\nonumber \\&-&\frac{P_{\alpha\beta}({\bf q}')}{\eta
{q'}^2} b^2 \xi_f f_\beta.
\end{eqnarray}
We choose $\xi_v=b^2\xi_f = b^{(d+z+2)/2}$. We also define $\alpha_0'=\alpha_0
\xi_v^{-1}\xi^2 b^{-(d+z+1)}$. This yields for the scaled coupling constant
\begin{equation}
 \mu'=\frac{\lambda'\alpha_0'}{2\eta\kappa'\pi}=\xi^2
b^{-2d-2z}\frac{\lambda\alpha_0}{2\eta \kappa\pi}=\mu,
\end{equation}
establishing the marginality of $\mu$ to $O(D/\kappa)^0$.

\section{Model III: NLS coupled with multiplicative noises}
\label{model3}

Notice that although Model I and Model II above do display  
phase
transitions for proper choices of the control parameters, the magnetic order
parameter remains zero on both sides of the transition point and there is no
true LRO in TL at $2d$. This sets them apart from the usual second order 
magnetic transitions and raises the question what
should be a minimal generalized NLS with additional degrees of freedom that may
show true
LRO in TL at $2d$.  True LRO in TL requires
finite $\Delta_0$ [see (\ref{fluctuations_pi}) above] in TL at $2d$ in the
stiff phase. In the stiff phases of Model I and Model II, $\kappa_e(q)\sim -\ln
(q a_0)$ in the long wavelength limit that is responsible for the $\ln\ln L
$ behavior of $\Delta_0$. In order to have a finite $\Delta_0$ in TL,
$\kappa_e(q)$ should diverge more strongly than $-\ln (q a_0)$ in TL at $2d$.
It is reasonable to expect that the presence of long range noises in the
system may lead to a finite $\Delta_0$ in TL at $2d$, indicating the
existence of true LRO in TL. We study this in this Section by
using  simple reduced models.

In this Section, we construct simplest possible generalizations of NLS by
minimally coupling NLS with multiplicative noises of given structures. It
remains to be seen how the explicit forms of the couplings with the
multiplicative noises affect the long wavelength properties. Consider first the
simple case of a scalar multiplicative noise $\hat g({\bf x},t)$, coupled to the
$O(N)$ spin $\Phi$ via symmetry-allowed minimal couplings. The nonconserved
relaxational dynamics of $\Phi$ is given by
\begin{equation}
 \frac{\partial \Phi}{\partial t}+\overline\lambda \hat
g\Phi=-\Gamma\frac{\delta
{\mathcal F}_\sigma}{\delta\Phi} + {\boldsymbol\theta},
\end{equation}
where $\overline\lambda$ is a coupling constant and $\hat g$ is a zero-mean 
 $\delta$-correlated Gaussian white noise with a given variance, say
$1$;
$\boldsymbol\theta$ is a zero-mean Gaussian white noise with a variance given by
(\ref{thetavari}). As before we use the parameterization
$\Phi=(\sigma,{\boldsymbol\pi})$ and for an assumed nearly ordered state,
$\sigma\approx 1$. The corresponding generating functional in a compact form is
\begin{eqnarray}
 {\mathcal Z}_0&=&\int {\mathcal D}\Phi{\mathcal D}\hat\Phi \exp[\int d^dx
dt \frac{D}{\Gamma}\hat\Phi\cdot\hat\Phi \nonumber \\&-&\int
d^dxdt\hat\Phi\cdot\{\frac{1}{\Gamma}\frac{\partial\Phi}{\partial
t}+\frac{\overline\lambda}{\Gamma} g\Phi + \frac{\delta {\mathcal
F}_\sigma}{\delta\Phi}\}].\label{z0}
\end{eqnarray}
Now as before, impose $\hat\Phi\cdot\Phi=0$. This evidently eliminates the
multiplicative noise term in (\ref{z0}), leaving it identical to that for pure
NLS. Thus, the ensuing dynamics is identical to that of pure NLS with $\hat g$
having no effects on it.

We now generalize $\hat g$ to vector multiplicative noises and study both the
nonconserved and conserved dynamics of NLS,
coupled to vector multiplicative noises $a_\alpha$ (with $d$-components in a 
$d$-dimensional space) via minimal couplings. The corresponding nonconserved
and
conserved equations of motion of NLS are
\begin{eqnarray}
 \frac{\partial\Phi}{\partial t}+\lambda_1 ({\bf a}\cdot{\boldsymbol
\nabla})\Phi &=& -\Gamma\frac{\delta {\mathcal
F}_\sigma}{\delta\Phi}+{\boldsymbol\theta},\label{nlsnoncon3}
\end{eqnarray}
and
\begin{equation}
 \frac{\partial\Phi}{\partial t}+\lambda_2{\boldsymbol\nabla}\cdot ({\bf
a}\Phi)=\Gamma\nabla^2\frac{\delta {\mathcal F}_\sigma}{\delta\Phi} +
\nabla_\alpha {\boldsymbol {\Theta}}_{\alpha},\label{nlscon3}
\end{equation}
respectively. Stochastic noise $\theta_i$ is Gaussian-distributed with zero
mean and variance as given by (\ref{thetavari}), where as stochastic noise
$\Theta_{i\alpha}$, again assumed to be zero mean Gaussian distributed, has a
variance
\begin{equation}
 \langle \Theta_{i\alpha}({\bf x},t)\Theta_{j\beta}({0},0)\rangle=2D\Gamma\delta({\bf
x})\delta (t)\delta_{ij}\delta_{\alpha\beta}.
\end{equation}
In addition, $a_\alpha$ is a vector (of $d$-components in a $d$-dimensional
space)
multiplicative noise. We choose $a_\alpha$ to be Gaussian-distributed with zero
mean and a variance
\begin{equation}
 \langle a_\alpha({\bf q},t) a_\beta({\bf -q},0)\rangle =
[mP_{\alpha\beta}+nQ_{\alpha\beta}]\frac{\delta (t)}{q^y},\label{multcons}
\end{equation}
in the Fourier space.
Here,
$Q_{\alpha \beta}= q_\alpha q_\beta/q^2=\delta_{\alpha \beta}-P_{\alpha\beta}$ is the longitudinal
projection
operator in the Fourier space, $P_{\alpha\beta}$ is the transverse projection
operator defined before. Positivity of $\langle |a_\alpha ({\bf
q},t)|^2\rangle$ implies $m(d-1)+n>0$, allowing anyone of $m$ or $n$ to become
negative, subject to the positivity of $\langle |a_\alpha ({\bf
q},t)|^2\rangle$. We choose $y$, a positive or negative
integer, as a free
parameter that defines the scaling of (\ref{multcons}). The presence of a
long-range noise coupled to the $O(N)$ may  be realized in a lattice-gas
type
model as follows. Imagine a $2d$ lattice, where sites are occupied by
(point) particles, with a spin $\Phi$ attached to it. Now assume that the
particles are
nonconserved (due, e.g., to evaporation and absorption) - hence their density
fluctuations are short-lived and drop out of the
dynamical descriptions of the model in the long-wavelength, long-time limit.
Now if
the particles are subject to Gaussian-distributed random
forces (e.g., if the particles are charged and encounter random electromagnetic
pulses, but neglect the inter-particle interactions) with appropriately chosen
variances, effective dynamical
equations for the $O(N)$ spins in this lattice-gas model in the long
wavelength, long time limit should have the form (\ref{nlsnoncon3}) on 
symmetry grounds, as the
total spin, being attached to the particles, is not conserved. On the other
hand, if there are no particle non-conserving events, but the diffusivities of the
particles are diverging, then again the density fluctuations relax fast
(keeping the spins conserved) and the effective equations of motion for the spin
variables should be of the structure as in
(\ref{nlscon3}). Notice that we do not specify any particular spin-particle
interactions above, as our arguments are sufficiently general and should hold
for any short range interactions between the spins and the 
particles~\cite{new-comment}.
 While this
is a simplistic motivation and  Model III is
admittedly artificially constructed, it  is a reduced model that is purposefully
designed to study (i) under what conditions,  true LRO might set
in NLS with additional degrees of freedom in TL at $2d$,
 and
(ii) whether the nonconserved and conserved dynamics yield the
same result in the long-wavelength limit. We will see below that
our results from Model III are sensitive to the value of $y$, such that
$\Delta_0$ may become independent of $L$ leading to LRO in $2d$ in the
nonconserved version of Model III, but not in its conserved version.

First consider the nonconserved version of Model III. The action functional
after expanding  up to the linear order in $D/\kappa$  is (see $S_I$ and
$S_{II}$  respectively for Model I and Model II above)
\begin{eqnarray}
 S_{IIInon}&=&\int d^dx dt \frac{D}{\Gamma} \hat\pi_i\hat\pi_i-\int d^dx dt
\hat\pi_i [\frac{1}{\Gamma}\partial_t\pi_i \nonumber \\&-&\kappa\nabla^2\pi_i]+\int d^dx dt
\frac{D}{\Gamma} (\hat\pi_j \pi_j)^2 \nonumber \\&+& \int d^dx dt \frac{1}{2} \hat \pi_i \pi_i
[-\frac{1}{\Gamma} \frac{\partial}{\partial t} \pi^2 \nonumber \\  &+&\kappa
\nabla^2 \pi^2 -2 h_1(1+ \frac{\pi^2}{2})]\nonumber \\ &-&
\frac{\lambda_1}{\Gamma} \int d^dx dt \hat \pi_m ({\bf
a}\cdot{\boldsymbol\nabla}) \pi_m \nonumber \\ &+& \rho \int
d^dx dt \pi^2.
\end{eqnarray}
If $\lambda_{1s}=\lambda_1/\Gamma=0$, one recovers the usual NLS action. Non-zero $\lambda_{1s}$
contributes only to $D/\Gamma$ to the leading order at $O(\lambda_{1s}^2)$, viz.,
$\frac{\lambda_{1s}^2}{2} \hat \pi_m \hat \pi_m \langle(a.\nabla)\pi_n
(a.\nabla) \pi_n\rangle$. This correction evaluates to
$(n \lambda_{1s}^{2}/2)\int \frac{d^dq}{(2\pi)^d}D/\kappa q^y$. Hence,

\begin{equation}
 \left( \frac{D}{\Gamma} \right)^{\prime} = \hat \xi^2 b^{-(d+z)} \frac{D}{\Gamma}[1+\Delta+\frac{n\lambda_{1s}^2 \Gamma}{2 D} \overline{\Delta}],
\end{equation}
where, $\overline{\Delta}=\int \frac{d^dq}{(2\pi)^d}D/(\kappa q^y)$.
Proceeding identically as in Model I and Model II, we obtain
for $D^{\prime}$:
\begin{equation}
 D^{\prime} = D b^z \hat \xi \xi^{-1} [1+ \frac{n\lambda_{1s}^2 \Gamma}{2 D}\overline{\Delta}].
\end{equation}
On imposing $D^\prime = D$,
\begin{equation}
 \hat \xi =\xi b^{-z}[1-\frac{n\lambda_{1s}^2 \Gamma}{2 D}\overline{\Delta}]
\end{equation}
Now, $\kappa^{\prime}=\kappa \hat \xi \xi b^{-(d+z+2)}(1+\Delta)$. On substituting $\xi=b^{d+z}[1- \frac{N-1}{2} \Delta]$,
\begin{equation}
 \kappa^{\prime}=\kappa b^{d-2} [1- (N-2)\Delta - \frac{n\lambda_{1s}^2
\Gamma}{2 D}\overline{\Delta}].\label{kappadisc}
\end{equation}
Notice that there is a contribution to $D^\prime$ with coefficient $n$
that originates from the multiplicative noise. Its evaluation in details is
given in Appendix~\ref{appendiag}.

We now have three distinct possibilities, namely $y>2, y<2$ and $y=2$.
Clearly, for $y>2$, for small $q$, $\overline{\Delta}$ dominates over $\Delta$,
where as, for $y<2$, $\overline\Delta$ is subdominant. Both terms are
equally relevant (in a scaling sense) for $y=2$. It is important to consider
the scaling of the parameter $\chi=\frac{n\lambda_{1s}^2 \Gamma}{2 D}$, that
appears as an effective expansion parameter (see below), for an arbitrary $y$.
To find that we look at the scaling of $\lambda_{1s}$ for
$y \neq 2$. Similar to Eq.~(\ref{lam_sc}), we can write
\begin{equation}
 \lambda_{1s}^{\prime}=\lambda_{1s}b^{-2d-2z-1}\hat \xi \xi \xi_a,
\end{equation}
where, $\xi_a$ is the rescaling factor for the field ${\bf a}$.  Using
Eq.~(\ref{multcons}), we evaluate
$\xi_a$:
\begin{equation}
 \xi_a=b^{y/2}b^{(d+z)/2}.
\end{equation}
Using values of $\hat \xi, \xi$ and $\xi_a$, we obtain
\begin{equation}
 \lambda_{1s}^{\prime}=\lambda_{1s}b^{\frac{d-z+y-2}{2}}
[1-(N-1)\Delta-\frac{n\lambda^2\Gamma}{2 D}\overline{\Delta}].
\end{equation}
Thus, up to order $(D/\kappa)^0$, $\lambda_{1s}^{\prime 2}=\lambda_{1s}^2
b^{d-z+y-2}$. Hence,
$\lambda_{1s}^{\prime 2} \Gamma^{\prime}=\lambda_{1s}^2 \Gamma b^{y-2} \implies
\chi^{\prime}=\chi b^{y-2}$. Thus, with $b=e^l$ and small $l$
\begin{equation}
 \frac{d\chi}{dl}=\chi (y-2).\label{chiflow}
\end{equation}
Therefore, $\chi$ is marginal for $y=2$, but grows (decays)
indefinitely for $y>(<) 2$.
Let us first consider the case
when $y=2$. Using $\epsilon=d-2$
\begin{equation}
 \frac{d\kappa}{d l}=\kappa [\epsilon - \frac{(N-2 +\chi)D}{2 \pi \kappa}].
\end{equation}
As long as $N-2+\chi >0$, FP
for the flow
equation for $\kappa$ yields a critical value $\tilde D^*$ for the reduced
noise strength $\tilde D\equiv D/\kappa$. We get
\begin{equation}
\tilde D^*=\frac{2\pi\epsilon}{N-2+\chi},\label{dmodel3}
\end{equation}
 demarcating
a ``high noise'' paramagnetic disorder phase and a ``low noise''
ferromagnetically ordered phase at $d=2+\epsilon$. At $d=2,\epsilon=0$, $\tilde
D^*=0$. This then implies the
absence of any ferromagnetically ordered state, except for
at
$\tilde D=0$, analog of $T=0$ for equilibrium systems. This again shows
that $d_L=2$.

\begin{figure}[htb]
 \includegraphics[height=6cm]{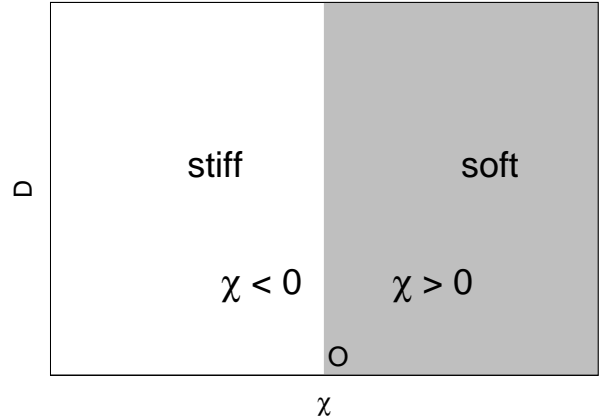}
 \caption{Phase diagram of nonconserved Model III in the $\chi-D$ plane at
$2d$ for $N=2$, $y=2$. Symbol $O=(0,0)$ is the origin. Stiff ($\chi<0$, 
renormalized $D/\kappa\rightarrow 0$) and 
soft with SRO
($\chi>0$, renormalized $D/\kappa$ diverges) are marked.}
 \label{dchi}
\end{figure}

Notice that negative values are allowed for $n$, hence $\chi$ can also be
negative. Focusing on $2d$ and with $\chi_c=-(N-2)$ we write
\begin{equation}
 \frac{d \kappa}{d
l}=\frac{(\chi_c-\chi)D}{2\pi}=-\frac{|\chi_c|+\chi}{2\pi}D\equiv
\frac{\Delta\chi D}{2\pi},
\end{equation}
where $\Delta\chi=-(|\chi_c|+\chi)$ giving
\begin{equation}
 \kappa=-\frac{(\Delta\chi)D}{2\pi}{\rm ln}(q
a_0)+\kappa_0,\label{kappanoncons3}
\end{equation}
where $\kappa_0$ is a microscopic stiffness. For $\chi<0$ and
$|\chi|>|\chi_c|$, $\Delta\chi>0$. Hence,
 $\kappa(q)$ diverges logarithmically in TL, $q \rightarrow
0$.
 Hence, the variance of the transverse spin
fluctuations
$\Delta_0 \sim \frac{1}{\Delta\chi}{ \ln \ln}L$.
This is reminiscent of our results in the stiff (ordered) phases of Model I and
Model
II, with $\chi$ playing the role of the control parameter. For
$\Delta\chi<0$, renormalized $\kappa$ vanishes for a large enough system
size, allowing us to define a persistence length $\zeta$ in direct analogy with
Model I and II
above. Similar to Model II, we generally find that $\zeta
(\chi>0)<\zeta(NLS)<\zeta(\chi<0)$. The phase transition
from soft to stiff phase may be described by the same order parameter as for
Model I and Model II. The flow of the reduced noise strength $\tilde
D=D/\kappa$
may be obtained
as in Model I and Model II. We find
\begin{equation}
 \frac{d\tilde D}{dl}=-\tilde D^2\frac{\Delta\chi}{2\pi},
\end{equation}
suggesting that for $\Delta\chi>0$, $\tilde D$ flows to zero in TL, implying an
ordered state, where as for $\Delta\chi <0$, $\tilde D\rightarrow\infty$ as
$L\rightarrow\zeta$, indicating a disordered state. This is similar to our
analysis for Model I. Notice that $\chi_c=0$ for $N=2$. Then following our
analysis for Model I, the Debye-Waller factor $\exp(-2W)$ and the equal-time
spin-spin
correlator $C_s(r)$ for $N=2$ at $2d$ may be calculated in the stiff phase at
$2d$. Not surprisingly and similar to the stiff phases in Model I and Model II,
we find $\exp(-2W)\rightarrow 0$ for $L\rightarrow\infty$, precluding any LRO.
Furthermore, $C_s(r)\sim 1/[\ln r]^{|\chi|},\chi <0$ in the stiff phase at $2d$.
Again not surprisingly, the spatial decay of $C_s(r)$ is the same as in the
stiff phases of Model I and Model II, and characterized by $|\chi|$, a
model-dependent quantity. A schematic phase diagram of the nonconserved version
of Model III in the $\chi-D$ plane is shown in Fig.~\ref{dchi} ($d=2,N=2$).

What happens when $y\neq 2$? Assume $y>2$; the $\chi$-term in (\ref{kappadisc})
clearly dominates.
Noting that $\int_{\Lambda/b}^\Lambda d^2q /q^y = l/\Lambda^{y-2},\,b=e^l\approx
1+l$, for $\chi <0$,
the recursion relation for $\kappa$ now reads
\begin{eqnarray}
 \frac{d \kappa}{d l}&=&\kappa \left[\epsilon+\frac{ D |\chi (l)|}{2\pi \Lambda^{y-2}
\kappa} \right],\label{eqmod3}
\end{eqnarray}
where $\epsilon=d-2$. Note that the pure NLS term has been dropped in 
(\ref{eqmod3})
since it is less infra-red divergent than the term arriving from
the noise for $y>2$.
Now at $2d$ setting $\epsilon=0$, $d\kappa/dl>0$,
implying $d_L<2$. Solving
Eq.~(\ref{chiflow}) and Eq.~(\ref{eqmod3}) together we obtain,
\begin{eqnarray}\label{kappchinon}
 \kappa (l) &=&\frac{|\chi| D}{2 \pi(y-2) \Lambda^{y-2}}(e^{l})^{y-2} +\kappa_0\nonumber \\
&\implies& \kappa(q)=\frac{|\chi| D}{2 \pi(y-2)} q^{2-y}+\kappa_0,
\end{eqnarray}
where $\kappa_0$ is a microscopic stiffness. Therefore, $\kappa(q)$ diverges as
a power law in $q$ as $q \rightarrow 0$.
It should be noted that for $\chi>0$, the recursion relation for $\kappa$ has the form
\begin{equation}
 \frac{d \kappa}{d l}=\kappa[\epsilon-\frac{ D \chi (l)}{2\pi \Lambda^{y-2}
\kappa}].
\end{equation}
Hence the $q$-dependence of $\kappa$ for $\chi>0$ is given by
\begin{equation} \label{forpers}
 \kappa(q)=\frac{-\chi D}{2 \pi(y-2)} q^{2-y}+\kappa_0
\end{equation}

Now for $\chi<0$, renormalized variance, $\Delta_0$ is given by
\begin{eqnarray}
 \Delta_0&=& \int_{L^{-1}}^{\Lambda} \frac{d q}{2 \pi} \frac{D}{q
\kappa(q)} = \frac{|\chi|D}{2\pi}\int_{L^{-1}}^{\Lambda} \frac{d q}{2 \pi}
\frac{1}{q^{3-y}}\nonumber \\&=&\frac{|\chi|D}{2\pi (y-2)}
\left[\frac{1}{q^{2-y}} \right]_{L^{-1}}^{\Lambda},
\end{eqnarray}
which gives a finite value for $\Delta_0$ in TL, $L\rightarrow\infty$,
indicating true LRO.
 Consider the Debye Waller factor $\exp(-2W)$, as defined above, and the
spin-spin correlation
$C_s(r)$ at $2d$ for $N=2$ with $\chi<0$. Proceeding in analogy with Model I,
we find
that $W (=\frac{\Delta_0}{2})$ is  finite in TL. Thus the order parameter does not vanish in
TL, indicating LRO in TL. In the same way,
\begin{equation}
 g({\bf x})=\frac{1}{2}\langle [\tilde\pi ({\bf x},t)-\tilde\pi
(0,t)]^2\rangle=\Delta_0 (L\rightarrow\infty),
\end{equation}
a finite number in the limit $r\rightarrow\infty$. Thus,
$C_s(r\rightarrow\infty)$ is
finite, consistent with a finite $W$ as obtained above. This firmly establishes
LRO (ferromagnetic order) for $\chi<0, y>2$ at $2d$.
 Notice that for
any
$y>2$ and $\chi <0$, the system is always in the ordered phase; there is no
phase transition at any $d$. This is because
renormalized $D/\kappa\rightarrow 0$ in TL at any
$d$ for $y>2,\chi <0$. The existence of LRO should not come
as a surprise, since nonconserved Model III is driven out of equilibrium by the
long-range multiplicative noise.  
We now briefly consider the consequence of $\chi>0$. Clearly, from
(\ref{forpers}), $\kappa(l)$ rapidly drops to zero as the system size
approaches the persistence length $\zeta$ given by
\begin{equation}
 \zeta\sim \left[\frac{\kappa_0(y-2)}{\chi D}\right]^{1/(y-2)}.
\end{equation}
Thus, nonconserved Model III with $y>2$ 
forms an example of {\em dynamic second order phase transition} 
between a disordered state with SRO and an ordered state with LRO 
with $\chi$ as a tuning parameter.  In dynamic phase 
transitions, a model goes from one phase to another upon tuning a model 
parameter of dynamic origin. Well-known examples of dynamic phase transitions 
include the Kardar-Parisi-Zhang (KPZ) equation for surface growth at dimensions 
$d>2$~\cite{kpz} and active-to-absorbing state phase 
transitions~\cite{hinrichsen}.  Our nonconserved Model III with $y>2$  
is an analog of these examples where a continuous symmetry is {\em dynamically 
broken}.

\begin{figure}[htb]
 \includegraphics[height=6cm]{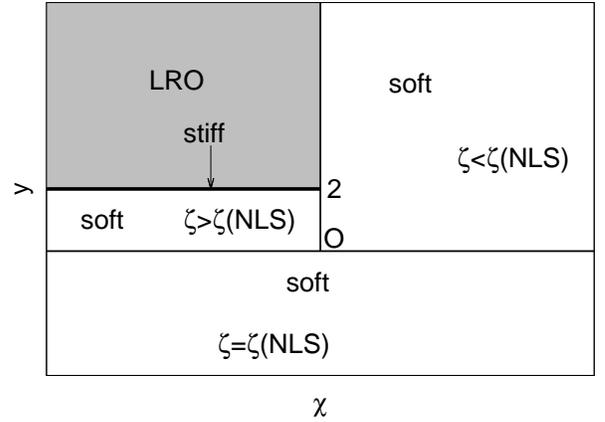}
 \caption{Phase diagram of nonconserved Model III in $\chi-y$ plane at $2d$
with $N=2$. Symbol $O$ denotes the origin (0,0). LRO, soft and stiff phases are 
marked.}
 \label{ychi}
\end{figure}

Lastly, for $y<2$, the pure NLS contribution in (\ref{kappadisc}) dominates in
the long wavelength limit, regardless of the sign of $\chi$. Thus, in that
limit the properties of
nonconserved Model III with $y<2$ is identical to that of pure NLS, with no
order at a finite $T$ in $2d$. A phase diagram of the nonconserved version of
Model III in the $\chi-y$ plane is shown in Fig.~\ref{ychi} with $d=2,N=2$.
Below in
Fig.~\ref{yepsi} we provide a phase diagram of nonconserved Model III in the
$y-\epsilon$ plane with $\chi<0$.
\begin{figure}[htb]
 \includegraphics[height=6cm]{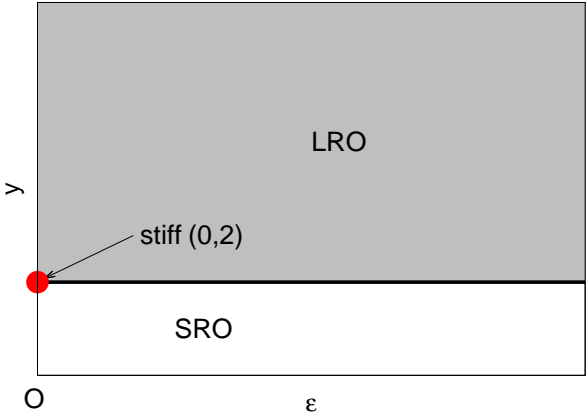}
\caption{(color online)Phase diagram of nonconserved Model III in $y-\epsilon$ 
plane. Regions
in the phase space with LRO, SRO (soft phase) and stiff phase (circle) are 
marked.}\label{yepsi}
\end{figure}

We now calculate the dynamic exponent $z$ at the DRG FP and in the 
stiff/LRO phases. For
$y<2$, results of the pure NLS ensue, precluding any stiff phase at $2d$ for all $N\geq 2$.
On the other hand, for $y>2$, $\chi<0$ 
we have for the recursion relation for $\Gamma$
\begin{equation}
 \Gamma^{\prime}=\Gamma b^{z-2-\epsilon} [1- |\chi| \overline{\Delta}].
\end{equation}
The recursion relation for $\Gamma$ for $y>2$, thus takes the following form:
\begin{equation}
 \Gamma^\prime = \Gamma b^{z-2-\epsilon} \left[1- \frac{|\chi| D (1-b^{y-2})}
 {2\pi (2-y)\Lambda^{y-2} \kappa} \right].
 \label{gammadis}
\end{equation}
As before, the dynamic exponent $z$ is obtained from the DRG flow equation for 
$\kappa\Gamma$. Considering $y>2$, using (\ref{eqmod3}) and proceeding as for 
Model I and Model II, we find $z=2$ in the LRO phase.
Precisely at $y=2$, again proceeding as for Model I and Model II, we have 
for $\chi < 0$
\begin{equation}
 z= 2 + \epsilon- \frac{(N-2-|\chi|) D}{2 \pi \kappa}
\end{equation}
at the unstable DRG FP for $|\chi| <|\chi_c|$. However, for $|\chi|>|\chi_c|$, 
$z=2$ in the stiff phase, as in Model I and Model II.

We now study the conserved version of Model III. The corresponding action, 
$S_{III_{con}}$ is given by
\begin{eqnarray}
 S_{III_{con}}&=&\int d^dx dt \frac{D}{\Gamma \nabla^4} (\nabla_{\alpha}\hat\pi_i)
 (\nabla_{\alpha}\hat\pi_i)\nonumber \\&-&\int d^dx dt
\hat\pi_i [-\frac{1}{\Gamma \nabla^2}\partial_t\pi_i -\kappa\nabla^2\pi_i]\nonumber \\&+&\int d^dx dt
\frac{D}{\Gamma \nabla^4} (\nabla(\hat\pi_j \pi_j))^2 \nonumber \\
&+& \int d^dx dt \frac{1}{2} \hat \pi_i \pi_i
[\frac{1}{\Gamma \nabla^2} \frac{d}{dt} \pi^2 +\kappa \nabla^2 \pi^2 -
2 h_1(1+ \frac{\pi^2}{2})]\nonumber \\&-&
\frac{\lambda_2}{\Gamma} \int d^dx dt \frac{\pi_m}{\nabla^2}
({\bf a}\cdot{\boldsymbol\nabla}) \hat\pi_m + \rho \int d^dx dt \pi^2
\end{eqnarray}
The leading contributions to $D/\Gamma$ come from (i)
$\frac{D}{\Gamma \nabla^4} \nabla_\alpha (\hat \pi_i \pi_i) \nabla_\alpha (\hat
\pi_j \pi_j)$ and  (ii) a correction second order in $\lambda_2/\Gamma$, viz.,
$\frac{\lambda_2^2}{2\Gamma^2}\frac{(\nabla_\alpha \hat \pi_i)(\nabla_\beta \hat
\pi_j)} {\nabla^2}\frac{<a_\alpha \pi_i a_\beta \pi_j>}{\nabla^2}$; see
Appendix~\ref{remodelIII}.

For the conserved model, the correlator,
$\langle\pi_i \pi_j\rangle=\int \frac{d^dq}{(2\pi)^d} \frac{d\Omega}{2\pi}
\frac{2D \delta_{ij}}{\Gamma q^2(\frac{\Omega^2}{\Gamma^2 q^4}+\kappa^2 q^4)}=
\int \frac{d^dq}{(2\pi)^d} \frac{D}{\kappa q^2}\delta_{ij}=\Delta\delta_{ij}$ (for small $h_1$).
Furthermore, $\langle a_\alpha \pi_i a_\beta \pi_j\rangle$ evaluates to $\int
\frac{d^dq}{(2\pi)^d}\frac{D}{\kappa q^{y+2}}[m(d-1)+n]{\delta_{\alpha\beta}\delta_{ij} 
\over d}=\hat \Delta{\delta_{\alpha\beta}\delta_{ij} \over d}$.
Using these values, we find
\begin{eqnarray}
\left(\frac{D}{\Gamma}\right)^{<}&=&\frac{D}{\Gamma}[1+\Delta+\frac{\lambda_2^2}
 { 2\Gamma D}{\hat \Delta \over d}],\\
 \frac{1}{\Gamma^{<}}&=&\frac{1}{\Gamma}[1+\Delta],\\
 \kappa^{<}&=&\kappa[1+\Delta],\\
 h_1^{<}&=&h_1[1+\frac{N-1}{2}\Delta].
\end{eqnarray}
The contribution to $D^<$ from the multiplicative noise may still be
represented diagrammatically as in Appendix~\ref{appendiag}. However, its
explicit
expression is clearly different from the corresponding form for the
nonconserved Model III.
Using the same scaling procedure for $q,\Omega,\hat {\boldsymbol\pi}$ and
$\boldsymbol\pi$ as above, we can now write
$\xi=b^{d+z}[1-\frac{N-1}{2}\Delta]$. These can be used to write the following
re-scaled corrected values of the parameters:
\begin{equation}\label{dgam_con}
\left(\frac{D}{\Gamma}\right)^{\prime}=b^{-(d+z-2)}{\hat \xi}^2 \frac{D}{\Gamma}
[1+\Delta+\frac{\lambda_2^2}{2\Gamma D}{\hat \Delta \over d}],
\end{equation}
\begin{equation}\label{gam_con}
 \frac{1}{\Gamma^{\prime}}=b^{-(d+2z-2)}\hat \xi \xi \frac{1}{\Gamma}[1+\Delta],
\end{equation}
\begin{equation}\label{k_conserved}
 \kappa^{\prime}=b^{-(d+z+2)}\hat \xi \xi \kappa[1+\Delta].
\end{equation}
Using Eq.~(\ref{dgam_con}) and Eq.~(\ref{gam_con}) and imposing $D^{\prime}=D$
as
before, we obtain
\begin{equation}
 \xi=b^z \hat \xi [1+ P_1 {\hat \Delta \over d}],
\end{equation}
where, $P_1=\frac{\lambda_2^2}{2\Gamma D}$
Now using the values of $\hat \xi$ and $\xi$,
\begin{equation}
 \kappa^{\prime}=\kappa b^{d-2}[1-(N-2)\Delta-P_1 {\hat\Delta \over d}].\label{kappaprime}
\end{equation}

Using values of $\hat \xi, \xi$ and $\xi_a$, we obtain
\begin{equation}
 (\frac{\lambda_{2}}{\Gamma})^{\prime}=\frac{\lambda_{2}}{\Gamma}b^{\frac{d-z+y+2}{2}}
[1-(N-1)\Delta-P_1{\hat{\Delta} \over d}]
\end{equation}
Thus, up to order $(D/\kappa)^0$, ${(\frac{\lambda_{2}}{\Gamma})^\prime
}^2=\frac{\lambda_{2}}{\Gamma}^2
b^{d-z+y+2}$. Hence,
$(\frac{\lambda_{2}}{\Gamma})^{\prime 2}
\Gamma^{\prime}=(\frac{\lambda_{2}}{\Gamma})^2 \Gamma b^{y} \implies
P_1^{\prime}=P_1 b^{y}$. Thus, with $b=e^l$ and small $l$
\begin{equation}
 \frac{d P_1}{dl}= y P_1 \implies P_1(l)=P_1 \exp [y l].\label{P1flow}
\end{equation}
Therefore, $P_1$ is marginal for $y=0$, but grows (decays)
indefinitely for $y>(<) 2$.

\par
The second and third terms in the RHS of (\ref{kappaprime}) compete when
$y=0$. For $y<0$, the
pure NLS behavior follows while
for $y>0$, the nonequilibrium contribution dominates at small $q$.
For $d=2$, $\int_{\Lambda/b}^{\Lambda} \frac{d^2q}{2\pi q^{y+2}}= \frac{l}{2\pi \Lambda^y}$, where, $b=e^l=1+l$
and $\Lambda$ is an upper wavenumber cut-off. Similar to the non-conserved case 
(see above), here we find for $y>0$
\begin{equation}\label{kapp_con_recr}
 \frac{d \kappa}{d l}=\kappa
\left[\epsilon-P\frac{D}{2\pi\Lambda^y\kappa
}
\right],
\end{equation}
where $\epsilon=d-2$ and $P=\frac{[m(d-1)+n]}{d}P_1$. Hence
$dP/dl=\frac{[m(d-1)+n]}{d}dP_1/dl=yP$.
At $d=2$,
\begin{equation}
 \frac{d\kappa}{d l}=-\frac{P D}{2\pi  \Lambda^y}.
\end{equation}
Solving the flow equations for $P$ and $\kappa$ together,
\begin{eqnarray}
 \kappa(l)&=&-\frac{P D}{2 \pi y \Lambda^y} (e^{ l})^y +\kappa_0 \nonumber \\&\implies&
 \kappa (q) = -\frac{P D}{2 \pi y} q^{-y} +\kappa_0.
\end{eqnarray}
Thus as $q\rightarrow 0$, renormalised $\kappa (q)$ runs away rapidly to large
negative values, implying
the absence of any stiff phase in TL. The persistence
length, $\zeta$ at which $\kappa(q)$ vanishes is given by
\begin{equation}
 \zeta \sim  \left[\frac{\kappa_0 y}{P D} \right]^{1/y},
\end{equation}
decreasing rapidly as $y$ increases for fixed $\kappa_0, P$ and $D$; thus
$\kappa(l)$
rapidly vanishes to $0$ at scale $\sim\zeta$. Clearly,
$\zeta <\zeta(NLS)$ for a large enough $\kappa_0$.
For $y<0$, evidently the results of the pure NLS holds. A schematic phase 
diagram of the conserved version of Model III in the $y-P$ plane is shown in 
Fig.~\ref{yp} below.
\begin{figure}[htb]
 \includegraphics[height=6cm]{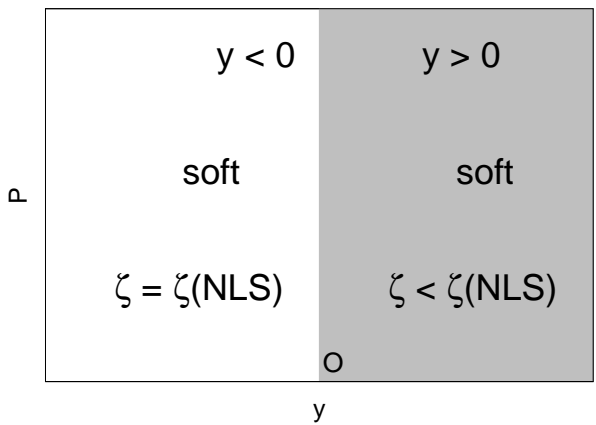}
 \caption{Phase diagram of nonconserved Model III in $y-P$ plane at $2d$ . Symbol, $O=(0,0)$
 denotes the origin. For both positive and negative values of $y$, the system remains
 soft but with different values of $\zeta$, as marked in the figure.}
 \label{yp}
\end{figure}

The take home message from this Section is that
merely the presence of long-range noise does
not automatically ensure phase transition to LRO in the system. The detailed
mechanism of coupling of the noise with the $O(N)$ and the underlying dynamics
(conserved versus nonconserved) as well as the structure of the noise 
(e.g., scalar
versus vector) are also equally important.
Significant differences between non-conserved and
conserved dynamics of Model III are not unexpected, as for
systems out of
equilibrium, the static or equal-time properties obtained from conserved and
nonconserved dynamics are not necessarily same, i.e., the equal-time properties
depend on the underlying dynamics. This is in contrast to the
well-known equivalence of ensembles in equilibrium systems.

\section{Summary and outlook}\label{conclu}
In this article, we have addressed phase transitions and the nature of order,
if any,
in certain generalized NLS constructed by us, where the $O(N)$ spins are
minimally coupled to
additional degrees of freedom. To this end,
we set up the dynamics of the nonconserved NLS with
(i) thermodynamically coupled Ising spins (Model I), (ii) a dynamically coupled
Stokesian velocity field (Model II) and (iii) a dynamically coupled vector
multiplicative noise (Model III). We also study the conserved version of Model
III. In
Model I,
the couplings are of thermodynamic origin, where as in Models II and III,
the couplings with the additional degrees are of dynamical origin. We use a
{\em low noise
variance expansion}, that for Model I is  identical to the well-known low-$T$
expansion for the pure NLS, where as for Model II and Model III, it
is a generalization of the standard low-$T$ expansion.
Our studies uncover {\em unusual phase transitions} in the models as
the relevant couplings exceed particular threshold or critical values;
in Models I and II, and in the nonconserved version of Model III for $y=2$,
the transition is between a {\em soft phase} with SRO, in which the renormalized 
spin
stiffness $\kappa$ vanishes in the long wavelength limit and a {\em stiff,
ordered
phase}, where $\kappa$ diverges as $\ln L$ in $2d$. In the latter phase, the
variance $\Delta_0$ of the transverse spin fluctuations scale proportional to
$\ln\ln L$ with a model-dependent proportionality constant at
$2d$, a dependence of $L$ that is substantially weaker
than the $\ln L$ dependence of the broken symmetry modes in classical elastic
Hamiltonians that exhibit QLRO. This implies a spatial decay of the equal-time
spin-spin correlation function in powers of inverse logarithm of the
(large) spatial separation, characterized by model-dependent exponents,
  at $2d$
for $N=2$. This model-dependence of the exponents are reminiscent of the 
model-dependent exponents that characterize the
algebraic decay of the spin-spin equal-time correlator in the equilibrium
XY model displaying QLRO; nevertheless, these power laws of inverse logarithm are
distinctly slower varying function of distance than the algebraic decay in QLRO.
Notice that this transition is {\em
not temperature driven}, but rather controlled by the coupling constants at
fixed $T$ or at the fixed noise amplitudes. These transitions have no
analog in pure NLS. Nonetheless,
 the magnetic order parameter remains zero even in the stiff
phase in TL at $2d$.
For $y>2$, $\Delta_0$ becomes {\em independent} of $L$ even at $2d$. This
implies LRO in the system. The presence of the long-range multiplicative noise
makes the model nonequilibrium and takes it out of the validity of MWT.
With $\chi$ as the tuning parameter, our nonconserved Model III with $y>2$ 
displays a dynamic second order phase transition.
We emphasize that there
is indeed no clash between our results and MWT. Notice however, the mere presence
of long-range noise itself does not necessarily lead to LRO, e.g., a
scalar multiplicative noise minimally coupled to the $O(N)$ spins have no effect and
the model is identical to pure NLS. In addition, the
conserved version of Model III does not display any LRO for
$y>0$. The lesson, therefore, is that the specific microscopic couplings 
between the noise and
the $O(N)$ spins
may lead to LRO. Furthermore, the conserved version of
Model III does not display any phase transition at $2d$, unlike its
nonconserved counterpart, and has long wavelength properties that are
qualitatively similar to pure NLS. This highlights the lack of {\em
equivalence} between the conserved and nonconserved dynamics of Model III,
emphasizing the underlying nonequilibrium nature of the dynamics. We also
obtained the dynamic exponent $z$ at the DRG FP and in the stiff phases of
Model I, Model II and nonconserved Model III, and in the LRO phase of
nonconserved Model III with long-range noise.

Our models here are admittedly artificially constructed and simple.
Nevertheless, they reveal important physical insight for more realistic but
complicated models. We note that in general, the additional degrees of freedom
can affect the dynamics of the $O(N)$ spins in two different ways: (i) through
new effective deterministic couplings for the $O(N)$ spins, and (ii)
modification of the noises, e.g., generation of effective noises in the
$O(N)$ spin dynamics. These features should be generic in NLS
coupled to other fields. Thus, the basic features of our results should
hold. While the Ising spins in Model I have independent dynamics, the overall
structure of the coupled dynamical equations is constrained by the FDT. 
Instead in Model II, the velocity field being assumed to be Stokesian, has no
independent dynamics, or, in Model III, the multiplicative noise is entirely
prescribed by its variance. An obvious generalization would be to consider
nonequilibrium coupling with additional fields having their own dynamical
evolutions, e.g., coupling with a growing surface described by a generalized
KPZ equation~\cite{kpz}. In this model, whether or not one
would find true LRO or an analog of the stiff phase, cannot
be immediately predicted. Lastly, we have entirely ignored the topological
defects. It is known that in $2d$, such defects lead to the Kosterlitz-Thouless
transition in the equilibrium XY model, where the system undergoes a transition from
low-$T$ QLRO
with bound pairs of defects to high-$T$ SRO with free defects~\cite{chaikin}.
It would be
interesting to find how the results of the present study may change if the
defects are accounted for in the models used here.
 We look forward to
detailed theoretical studies in this direction.

\section{Acknowledgement}
One of the authors (AB) gratefully acknowledges partial financial support by
the Max-Planck-Gesellschaft (Germany) and Indo-German Science \& Technology
Centre (India) through the Partner Group programme (2009).

\appendix
\section{Action for Model I}\label{remodel1}

We begin with the action functional $S_I$ for Model I. We perform the following rescalings: (i) rescale
$t\rightarrow t/\kappa$, (ii) absorb a factor $\sqrt {D/\kappa}$ in $\hat\pi_m$, (iii) absorb a factor of $\sqrt{\kappa/D}$
in $\pi_m$, (iv) $\psi\rightarrow \sqrt D\psi$,(v)$\hat\psi\rightarrow \hat\psi \kappa/\sqrt D$. This ensures
that the equal-time correlators $\langle\pi_m ({\bf q},t)\pi_m({\bf -q},t)\rangle$ and $\langle
\psi({\bf q},t)\psi({\bf -q},t)\rangle$ are independent of the model parameters, and hence $O(D/\kappa)^0$. We find
for the rescaled action
\begin{eqnarray}
S_I&=&\int d^dxdt
[\frac{1}{\Gamma}\hat\pi_m\hat\pi_m-\hat\pi_m\{\frac{1}{\Gamma}\partial_t
\pi_m
-\nabla^2\pi_m\nonumber \\&-&\frac{2\lambda D}{\kappa}\nabla_\beta(\psi^2\nabla_\beta \pi_m)\}
+\frac{1}{\Gamma}\frac{D}{\kappa}(\hat\pi_m\pi_m)^2\nonumber \\&+&\frac{1}{2}\frac{D}{\kappa}\hat\pi_m\pi_m
(-\frac{1}{\Gamma}\partial_t \pi^2
+\kappa\nabla^2\pi^2-\frac{h_1}{\kappa}\pi^2)\nonumber \\&-&\frac{h_1}{\kappa}\hat\pi_m\pi_m]
+\int d^dx dt\frac{1}{\kappa}\frac{D}{\kappa}\rho\pi^2 \nonumber \\&+&\int d^dx dt [\hat\psi\hat\psi\frac{1}{\Gamma_2'}
-\hat\psi(\partial_t\psi\frac{1}{\Gamma_2} +r\psi
-\nabla^2\psi)],\label{actionIrescale}
\end{eqnarray}
where $\Gamma_2'=\frac{D}{T \kappa}$, $\Gamma_2=1/\kappa$.
Clearly, {\em all} the nonlinear terms are $O(D/\kappa)$.
This justifies our perturbative expansion in small $D/\kappa$. Thus,
the bare propagator and correlators, which contain only terms that are
$O(T/\kappa)^0$, after Fourier transforming in space and time are
\begin{eqnarray}\label{prop pi}
 \langle \hat\pi_i\pi_j \rangle
&=&\frac{\delta_{ij}}{\frac{-i\omega}{\Gamma}+\kappa q^2+h_1},\\
 \langle \pi_i.\pi_j
\rangle&=&\frac{2D/\Gamma}{\frac{\omega^2}{\Gamma^2}+(\kappa q^2+h_1)^2}
\delta_{ij},\label{corr pi}\\
 \langle \hat\psi\psi \rangle&=&\frac{1}{-i\omega+r+q^2},\label{prop psi}\\
 \langle \psi\psi \rangle&=&\frac{2T}{\omega^2+(r+q^2)^2}.\label{corr psi}
\end{eqnarray}

\section{Action for Model II}\label{remodel2}

Use the same scaling for $\pi_m$ and $\hat\pi_m$ as in Sec.~\ref{remodel1}
above to obtain

\begin{eqnarray}
S_{II}&=&\int d^dxdt
[\frac{1}{\Gamma}\hat\pi_m\hat\pi_m-\hat\pi_m\{\frac{1}{\Gamma}\partial_t
\pi_m
-\nabla^2\pi_m\nonumber \\&-&\frac{2\lambda D}{\kappa}\nabla_\beta(\psi^2\nabla_\beta \pi_m)\}
+\frac{1}{\Gamma}\frac{D}{\kappa}(\hat\pi_m\pi_m)^2\nonumber \\&+&\frac{1}{2}\frac{D}{\kappa}\hat\pi_m\pi_m
(-\frac{1}{\Gamma}\partial_t \pi^2
+\kappa\nabla^2\pi^2\nonumber \\&-&(h_1/\kappa)\pi^2)-(h_1/\kappa)\hat\pi_m\pi_m]\nonumber
\\&+&\int d^dx dt\frac{1}{\kappa}\frac{D}{\kappa}\rho\pi^2 \nonumber \\&-& \lambda \alpha_0\int d^dx dt
(D/\kappa) [\hat \pi_i \frac{P_{\alpha \beta}}{\eta \nabla^2}
(\nabla_\beta \pi_j)(\nabla^2 \pi_j)\nabla_{\alpha} \pi_i],\nonumber \\\label{actionIIrescale}
\end{eqnarray}

\section{Action for Model III}\label{remodelIII}

Consider the nonconserved Model III with a vector multiplicative noise $\bf a$.
Averaging the generating functional over $\bf a$ leads to a term in
$S_{IIInon}$ of the form
\begin{eqnarray}
 &&\lambda_{1s}^2\int d^dx dt d^dx' dt'\hat\pi_m({\bf x},t)\hat\pi_n({\bf x'},t')
\nonumber\\&\times&\langle({\bf a}({\bf x},t)\cdot {\boldsymbol\nabla})\pi_m ({\bf
a}({\bf x'},t')\cdot{\boldsymbol\nabla})\pi_n ({\bf x'},t')\rangle_a,
\label{new}
\end{eqnarray}
where $\langle...\rangle_a$ implies averaging over the distribution of $\bf a$.
Noting that $\bf a$ is $\delta$-correlated in time and then using the same
scaling as in Sec.~\ref{remodel2} above, the new term (\ref{new}) above gets a
scale factor $(\lambda_{1s}^2/D)(D/\kappa)$. Averaging over $\bf a$ yields
additional factors containing $m$ and $n$; see (\ref{multcons}). This yields a
factor $\sim \chi D/\kappa$  for the contribution to $D$. All other terms in
$S_{IIInon}$ are same as those in the action for pure NLS. This establishes the
expansion of $S_{IIInon}$ to $O(D/\kappa)$. The action $S_{IIIcon}$ for
conserved Model III may similarly be expanded up to $O(D/\kappa)$.

The correlator and propagator terms for conserved Model III are given by:
\begin{eqnarray}
 <\pi_m \pi_n>=\frac{2D/\Gamma}{\frac{\omega^2}{\Gamma^2 q^4}+(\kappa q^2+h_1)^2} \delta_{mn}\\
 <\hat \pi_m \pi_n>=\frac{\delta_{mn}}{\frac{-i \omega}{\Gamma q^2}+\kappa q^2 +h_1}
\end{eqnarray}

\section{Alternative formulation for $N=2$}\label{O2}
In case of $O(2)$ spins, using $\Phi^2=1$, we write $\Phi=\exp [i \Psi]$. Here,
phase $\Psi$ is a Goldstone or broken symmetry mode.
Any variation in $\Phi$ is incorporated as a variation in the phase $\Psi$. The equations of motion in terms of $\Psi$ can thus be written as :

Model I :

\begin{equation}
 \frac{\partial \Psi}{\partial t}= \Gamma [\kappa \nabla^2 \Psi - 2\lambda
\nabla(\psi^2 \nabla \Psi)] +\Theta_{\Psi}.
\end{equation}

Model II :

\begin{equation}
 \frac{\partial \Psi}{\partial t} +\lambda {\boldsymbol v}. \nabla \Psi = \Gamma
\kappa \nabla^2 \Psi + \Theta_{\Psi}.
\end{equation}

Model III : Non conserved case

  \begin{equation}
    \frac{\partial \Psi}{\partial t} +\lambda_1 {\boldsymbol a}. \nabla \Psi =
\Gamma \kappa \nabla^2 \Psi + \Theta_{\Psi}.
  \end{equation}

Model III : Conserved case

\begin{equation}
  \frac{\partial \Psi}{\partial t} +\lambda_2 \nabla. ({\boldsymbol a}\Psi) =
\Gamma \nabla^2
[-\kappa (\nabla^2 \Psi)] + \nabla_{\alpha}\overline{\Theta}_{\Psi \alpha III}.
\end{equation}
Here, stochastic noises $\theta_\Psi$ and $\Theta_{\Psi i}$ are zero-mean
Gaussian white noises related to the noises $\boldsymbol\theta$ and
${\boldsymbol\Theta}_i$ above respectively. In each of the above cases,
renormalization of $\kappa$ may be calculated to the leading order in
$D/\kappa$, which are identical to the results obtained for Model I, Model II
and Model III above with $N=2$. Equal-time spin correlation functions $C_s(r)$
as defined above may be calculated in straight forward ways yielding same
results as above.

\section{Renormalization of $\kappa$ in Model
II}\label{noren}

The term $\int d^dx dt \lambda \hat \pi \frac{P_{\alpha \beta}}{\eta \nabla^2} f_\beta \nabla_\alpha \pi_i$,
in the second order evaluates to zero and hence does not contribute to $\frac{D}{\Gamma}$.

We now show that there is no renormalization to $\tilde\lambda$ at the one-loop 
order in Fig.~\ref{3point}:

\begin{figure}[htb]
\includegraphics[height=6cm]{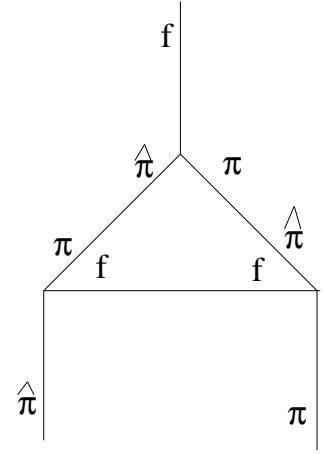}
\caption{Feynman diagram for one loop correction to $\lambda$ in Model II.} \label{3point}
\end{figure}

\begin{figure}[htb]
\includegraphics[height=6cm]{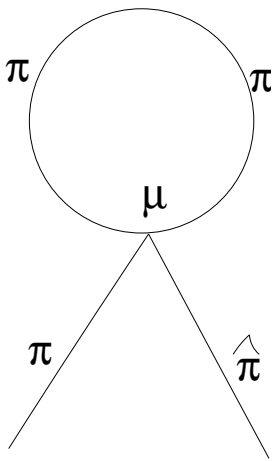}
\caption{Feynman diagram for $O(D/\kappa)$ correction in Model II.} \label{feedback}
\end{figure}

This evaluates to
\begin{eqnarray}
&\sim& \int d^dq d\Omega P_{\alpha\beta}({\bf q}) P_{\alpha\beta}({\bf 
q})\frac{\langle f_\gamma ({\bf q},\Omega)f_\delta({\bf 
-q},-\omega)\rangle}{\eta^2 q^4}\nonumber \\&\times&q_{1\mu}\langle\pi_\mu 
({\bf -q},-\Omega)\hat\pi_\nu({\bf q},\Omega)\rangle \langle\pi_\nu ({\bf 
-q},-\Omega)\hat\pi_\gamma({\bf q},\Omega)\rangle\nonumber \\
\end{eqnarray}
which vanishes due to causality. Now, we show that the coupling constant 
$\alpha_0$ has no correction to $O(D/\kappa)^0$. The relevant diagram is
shown in Fig.~(\ref{feedback}).

Fig.~(\ref{feedback}) gives the following
contribution to $\kappa$
\begin{eqnarray}
\sim \mu \int d^dq_1 d\Omega \hat \pi_i P_{\alpha \beta}({\bf q}_1) 
\langle|\pi_j ({\bf q-q}_1, \omega-\Omega)|^2\rangle,
\end{eqnarray}
as available in Section~\ref{two} above.

Clearly, the above diagram being proportional to $\langle 
\pi_\alpha\pi_\beta\rangle$ is proportional to $D/\kappa$. Thus, there are no 
corrections to $O(D/\kappa)^0$.


\section{Contribution of the multiplicative noises to one-loop corrections for
$D$ in Model III}\label{appendiag}

The one-loop diagram for the contributions upto $O(D/\kappa)$ to 
$\frac{D}{\Gamma}$ in the non-conserved 
version of 
Model III is shown in Fig.~\ref{model_nonfig}:

\begin{figure}[htb]
\includegraphics[height=3.5cm]{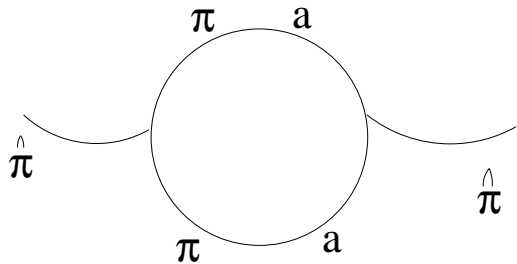}
\caption{Feynman diagram for the relevant correction in the non-conserved version of Model III.} \label{model_nonfig}
\end{figure}


\begin{thebibliography}{99}
\bibitem{lesstc} Fluctuations of $\psi$ are short lived even for $T<T_c$. 
However, one needs to keep the $u$-term in (\ref{scalar_free}) for thermodynamic 
stability. We do not discuss this here in details.
\bibitem{convolution}  Note that there is a convolution of the 
terms
$(\nabla_\alpha \pi_j) (\nabla^2 \pi_j)$ in the Fourier space which yields the 
scale
factor of $b^{-d-z}$ along
with $b^{-1}$ in Eq.~\ref{conv_scal}.
\bibitem{foot1} The absence of LRO at $2d$ even for
$\lambda>\lambda_c(\epsilon=0)$ is consistent with $d_L=2$.
\bibitem{foot2} While the order parameter $\overline O$ does distinguish 
the stiff phase from the soft phase, it is unusual as an order parameter in the 
sense that it is not defined {\em locally}.
\end{thebibliography}
\end{document}